\documentstyle{mn}
\input{epsf}

\begin{document}
\title[Evolution of the Hubble Sequence]
{Evolution of the Hubble Sequence in Hierarchical Models for Galaxy 
Formation}

\author[C.M. Baugh, S. Cole, C.S. Frenk]
{C.M. Baugh, S. Cole, C.S. Frenk. 
 \\
Department of Physics, Science Laboratories, South Road, Durham DH1 3LE}

\maketitle

\def\mpc {h^{-1} {\rm{Mpc}}}
\def\and  {\it {et al.} \rm}
\def\rmd {\rm d}

\begin{abstract}

We present a model for the broad morphological distinction between the disk
and spheroidal components of galaxies. Elaborating on the hierarchical
clustering scheme of galaxy formation proposed by Cole {\it et al}, we
assume that galaxies form stars quiescently in a disk until they are
disrupted into a spheroidal configuration by mergers.  Bulges and spheroids
may continue to accrete gas from their hot coronae, and so they may grow
disks again. Thus, an individual galaxy may pass through various phases of
disk or spheroid dominance during its lifetime. To distinguish between
disks and spheroids we add one additional free parameter to the
semianalytic model of Cole \and which we fix by requiring that the
predicted morphological mix should match that observed locally. Assuming an
$\Omega=1$, standard cold dark matter cosmology, we calculate formation and
merging histories, and the evolution in colour, luminosity and morphology
of the galaxy populations in different environments. Our model predicts
that the bulges of spirals were assembled before the spheroids of
ellipticals and the spheroids of cluster ellipticals were assembled before
those of field ellipticals. About 50\% of ellipticals, but only about $15 \%$
of spirals, have undergone a major merger during the redshift interval $0.0
\le z \le 0.5$. In spite of their violent formation history, elliptical
galaxies turn out to have colour-magnitude diagrams with remarkably small
scatter. Apart from a general blueing of the galaxy population with
redshift, the colour-magnitude diagrams are remarkably similar at redshift
$z=0.5$ and at the present day. 
The morphological mix of galaxies that become rich cluster members at high 
redshift is dominated by spiral galaxies, 
due to the long timescale for galaxy mergers compared with the timescale 
for cluster assembly at high redshift.
The assembly of low redshift clusters is slower, allowing more galaxy 
mergers to occur in the progenitor halos.
As a result $z=0$ rich clusters become E/S0 dominated and we find 
a ``Butcher-Oemler'' effect that becomes weaker for
poorer groups at high redshift. The field luminosity function of red
galaxies shows little evolution out to $z\simeq 1$ and the reddest galaxies
at these redshifts are as bright as their local counterparts. The blue
luminosity function, on the other hand, evolves rapidly with redshift,
increasing its characteristic luminosity and becoming steeper at the faint
end. These trends are similar to those recently observed in the
Canada-France Redshift Survey. Our calculations serve to demonstrate that a
simple prescription for the distinction between disks and spheroids that is
compatible with hierarchical clustering goes a long way towards explaining
many of the systematic trends observed in the galaxy population.

\end{abstract}


\section{Introduction}

The fundamental question of why some galaxies appear flattened like disks
while others are spheroidal remains a major unsolved problem in galaxy
formation. Attempts have been made in the past to present this problem as
a `nature' {\it versus} `nurture' dichotomy: is the morphology of a galaxy
imprinted in the initial conditions or does it result from environmental
processes such as mergers and tidal interactions? Sandage \and (1970)
postulated that the determining factor is the protogalactic angular
momentum: a disk or a spheroidal configuration would result, respectively, 
from the collapse of a rapidly rotating or of a slowly rotating gas cloud.
Building upon the classic study of the kinematics of stars in our galaxy
by Eggen, Lynden-Bell and Sandage (1962), Larson (1975) and 
Gott \& Thaun (1976) 
proposed that the significant parameter is the ratio of the star
formation timescale to the free-fall timescale. If the former is shorter,
dissipationless collapse would lead to a spheroidal system, while disks
would form from the dissipational settling of a rotating gas cloud. The
`nurture' hypothesis was first strongly advocated by Toomre (1977) who
argued that galaxies form initially as disks and ellipticals form
subsequently through violent mergers of disks.

A simple distinction between `nature' and `nurture' is not easy to
accommodate within modern cosmological theory. In hierarchical clustering
models, initial conditions and environmental effects are closely linked
together. Protogalactic objects are `born' with a range of angular momenta
and characteristic timescales and these as well as their environment 
and merger rates depend on the initial conditions. 
From an observational point of view, it is also clear 
that galaxy morphology reflects the interplay of various phenomena.
Dressler's (1980) morphology-density relation indicates that dynamical
processes that depend on environment are a major influence on the final
configuration of a stellar system. The well known excess of blue
galaxies in clusters at high redshift -- the Butcher-Oemler effect
(Butcher \& Oemler 1978, 1984; Allington-Smith \and 1993) -- suggests that
the predominance of certain morphological types depends also on cosmic
time. This impression seems to be supported by the recent HST images of
clusters at redshifts $z \sim 0.5$, which display an abnormally high
proportion of late spiral and irregular types (Dressler \and
1994, Couch \and 1994, Moore \and 1995).

Mechanisms that may transform one morphological type into another have
been proposed and investigated in the past two decades. These range from
ram pressure stripping of spirals by a hot intracluster medium (Gunn \&
Gott 1972), to galaxy `harassment' by impulsive encounters in clusters
(Moore \and 1995). Galaxy mergers continue to be a favourite explanation
for the origin of at least some bright ellipticals (Schweizer \& Seitzer 1992) and
for the predominance of ellipticals and S0's in rich clusters. N-body
simulations of merging disk galaxies do indeed produce merger remnants
with density profiles resembling those of ellipticals (eg Negroponte \&
White 1983). Nevertheless, objections have been raised against the view
that the majority of ellipticals result from the merger of two or more
stellar fragments. Perhaps the most powerful of these is the very small
scatter measured in the colour-magnitude diagram of rich cluster
ellipticals and S0's. For example, Bower \and (1992) find that the {\it
rms} scatter in the colour of elliptical galaxies in the Coma cluster is
only $\delta (U-V) < 0^{m}.04$. On this basis they argue that the bulk
of the stars in these galaxies are between 8 and 12 Gyr old and that
the galaxies themselves must have been assembled at that time. 

The origin of galaxy morphology is now beginning to be addressed with
numerical simulations that model both gravitational dynamics and gas
physics. Some studies have concentrated on achieving a detailed
understanding of mergers and interactions (For a review see Barnes \&
Hernquist 1992; some of the results of numerical simulations that are
relevant to this paper are discussed in Section \ref{s:bulge}) while
others attempt to follow the formation of galaxies in their proper
cosmological context (eg. Katz, Hernquist \& Weinberg 1992, Evrard,
Summers \& Davis 199?, Navarro, Frenk \& White 1995 and references
therein.) Although interesting results have emerged from these studies,
these are still early days in this area and a number of difficulties,
particularly regarding the behaviour of cooling gas and its transformation
into stars, require further study. In addition, these simulations are
expensive in computer time and so only a limited range of parameter space
has been explored to date.

An alternative approach to studying galaxy formation is seminanalytic
modelling. This has already proved to be a powerful technique (White \&
Frenk 1991, Cole 1991, Kauffmann \and 1993, Lacey \and 1993, Cole \and 1994). This
approach is based upon an extension of the Press-Schechter formalism
(Press \& Schechter 1974; Bond \and 1991, Bower 1991) whereby a
Monte-Carlo realisation of the hierarchical clustering process is used 
to follow the collapse and merger history of dark matter halos. The
current level of understanding of the dynamics of cooling gas, star
formation, feedback of energy into prestellar gas and galaxy mergers is
encoded into a few simple rules that allow the process of galaxy formation
within dark matter halos to be calculated. A wide range of parameters and
cosmologies are accessible within such models (eg Cole \and 1994, Heyl
\and 1995).

The semianalytic models have enjoyed a number of significant successes,
providing explanations for the general features of the galaxy luminosity
function, the slope and scatter of Tully-Fisher relation, the faint galaxy
number counts, their redshift and colour distributions, etc. However, a
number of details remain unresolved. For example, in the original Cole
\and (1994) model, galaxies as red as many observed ellipticals were not
formed. This problem turned out to be due to approximations in the Bruzual
and Charlot (1993) stellar population synthesis models used by Cole \and 
With the revised models of Charlot \and (1995), the colour distributions
are in much better agreement with observations (see Figure 2 of Frenk \and
1995). The strong feedback restricting star formation in low circular
velocity halos assumed by Cole \and (1994) resulted in a relatively flat
faint end slope for the luminosity function, but in spite of this, the
models still produce more dwarf galaxies than observed locally by Loveday
\and (1992). More recent determinations, however, indicate that the faint
end of the luminosity function is still uncertain (McGaugh 1994) and the
revised model of Cole \and in fact agrees quite well with the CfA
luminosity function of Marzke \and (1994; see Figure 3 of Frenk \and
1995). The Tully-Fisher relation recovered by Cole \and has a scatter and
slope that matches those observed, but is offset from the data. This
problem can be traced back to an overproduction of galactic size dark
halos in cold dark matter (CDM) cosmologies (Kauffmann \and 1993, Heyl
\and 1995).

In this paper, we present an extension of the model of Cole \and (1994)
that allows the light of each galaxy to be split into a disk and a bulge
component. The main aim of this paper is to use this model in order to
test the extent to which the distinction between disk and spheroidal
systems may be accounted for by the simplest scheme compatible with
hierarchical clustering. We shall assume (for the reasons given in
Section~2 ) that stars form initially in a disk and that bulges and
spheroids are formed from violent mergers of such disks. We therefore
ignore secondary dynamical processes such as dynamical friction or galaxy
harassment in clusters. With such a simplified scheme, we do not seek to
account for the details of the morphological sequence, but merely to
understand the basis of a broad morphological classification based upon
the bulge-to-disk luminosity ratio. 

Our study extends earlier analyses by Kauffman \and (1993) and Kauffmann
(1995b, 1995c) who also proposed a merger driven scheme for the
production of galactic bulges, with morphological classifications based on
bulge-to-disk ratios. They predicted the colours and morphological mix of
galaxies in clusters as a function of absolute magnitude and examined the
spiral or blue fraction and the colour-magnitude relation of galaxies in
high-redshift clusters.  Whilst the galaxy formation schemes of Kauffmann
\and (1993) and Cole \and (1994) are similar in spirit, there are
differences in detail in the way that the evolution of dark matter halos,
star formation, feedback and galaxy mergers are treated. In addition, there
are significant differences in our schemes for bulge formation which we
discuss in Section \ref{s:bulge}. Finally, in this paper we concentrate on
somewhat different issues from those investigated by Kauffman \and 

A brief outline of the galaxy formation scheme of Cole \and (1994) is given in
Section \ref{s:galform}. We describe the mechanism for bulge formation in
Sections \ref{s:bulge} and \ref{s:fellip}. Some representative examples of
galaxies with different morphologies are given in Section \ref{s:rep}. We
examine the systematics of bulge formation in Section \ref{s:bulgeform}.
The predictions of our model for the colours of galaxies as a function of
type and environment, the morphological mix in different environments and
the evolution of the luminosity function are compared with observations in
Section \ref{s:results}. Finally, we discuss the success and failings of
our model in Section \ref{s:discuss}.

A detailed comparison of the predictions of our model with the faint galaxy
counts as a function of morphological type obtained by the Hubble Space
Telescope Medium Deep Survey, is given in a separate paper (Baugh \and 1996).

\section{A Model for Galaxy Formation}

\begin{figure*}
{\epsfxsize=16.truecm \epsfysize=12.truecm 
\epsfbox[0 410 460 740]{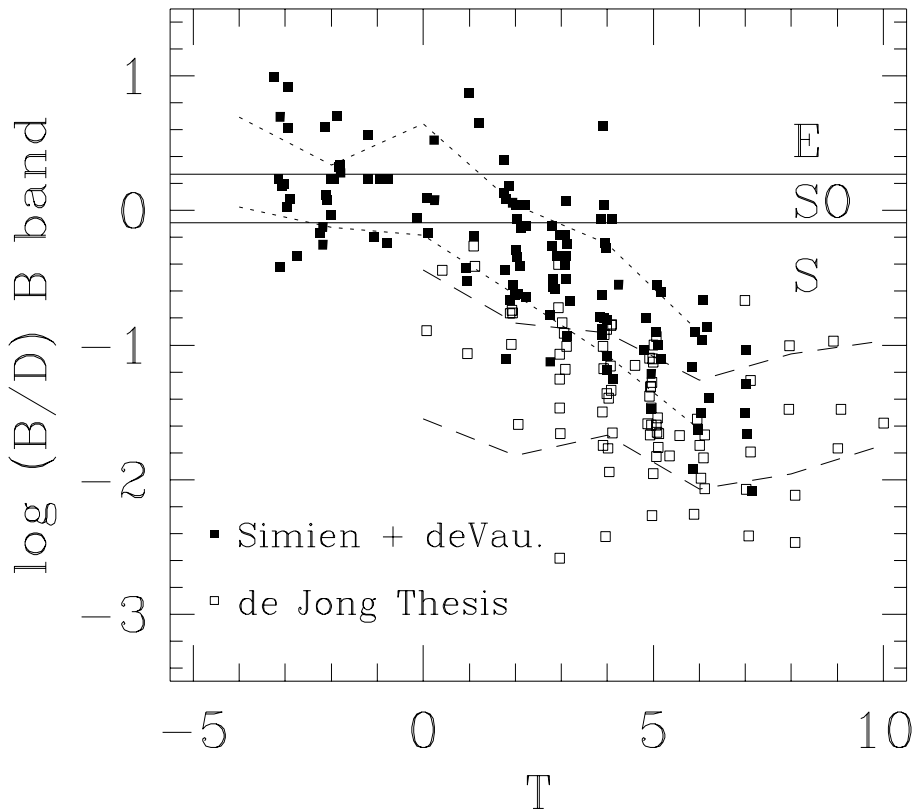}}
\caption[junk]{The correlation between bulge to disk ratio in the B band
and Hubble T type. The Simien and de Vacouleurs data is for a sample of 96
galaxies; the data from de Jong is for 86 face on spirals. The de Jong
data is for a 2D decomposition using exponential profiles for the bulge
and disk. The points have been given a small, random, horizontal
displacement for clarity. The dotted lines are the 20 and 80 percentiles
for the Simien \& de Vacouleurs data; the dashed lines are the 20 and 80
percentiles for the de Jong data. The two solid lines show the cuts on
bulge to disk ratio that we use in this paper to make a broad distinction
between morphological classes.
}
\label{fig:bddata}
\end{figure*}

\subsection{Galaxy Formation Scheme}
\label{s:galform}

In our semianalytic scheme, the complexities of galaxy formation are
approximated by a set of simple rules which, wherever possible, are
motivated by the results of numerical simulations.  Full details of the
basic scheme can be found in Cole \and (1994). Below we give a brief
outline of how the various relevant physical processes are incorporated
into the model.

The collapse and merging of dark matter halos as a function of time are
followed using the block model of Cole \& Kaiser (1988) (see also 
Cole 1991).  This provides
information about the distribution of halos as a function of mass, the
redshift at which the halos collapse, the redshift at which they merge into
bigger halos, and the properties of the halos into which subunits merge.
Each dark matter halo has baryonic material associated with it. When the
halo forms, the associated gas is shock heated and attains the virial
temperature of the halo.  Over the lifetime of the halo, defined as the
time interval between the formation of the halo and its merger with a
larger halo or the redshift of interest, some fraction of this hot gas can
cool, according to the standard cooling function for primordial abundances
of helium and hydrogen.

Once the gas has cooled it can form stars.  The star formation rate is
assumed to be proportional to the mass of cold gas present divided by a
star formation timescale.  The latter is independent of
the dynamical time at the epoch when the halo collapses, but is a function
of the circular velocity of the dark matter halo. This is because the
energy liberated by stellar winds and supernovae reheats some of the gas,
giving rise to a feedback loop involving cooling, star formation and
heating.  The details of this feedback mechanism are based upon the results
of `smooth particle hydrodynamics' simulations by Navarro \& White (1993)
and has a stronger dependence on halo circular velocity than is predicted
by a simple binding energy argument (White \& Rees 1978, White \& Frenk
1991).  Without feedback, the majority of gas in the universe would cool
at high redshift and form stars in low mass halos, precluding the
possibility of star formation in more massive halos later on (White \&
Rees 1978, White \& Frenk 1991, Cole 1991, Lacey \and 1993). Our feedback
prescription strongly inhibits the formation of stars in halos with low
circular velocity.

When a dark matter halo merges into a larger object, its hot gaseous corona
is stripped and the gas becomes associated with the new halo.
However, the cold gas and stars of the progenitor galaxies remain as
separate entities until the galaxies themselves merge. When a new halo is
formed, we compute a timescale for the merger of its galaxies, based upon
a formula again motivated by numerical simulations (Navarro, Frenk \& White
1995). This has a dependence upon the mass ratio of the participant
galaxies that is weaker than that suggested analytically from pure
dynamical friction considerations. If the galaxy merger timescale is
shorter than the lifetime of the halo, the galaxies merge at the epoch at
which the halo forms. The central galaxy in a halo accretes gas that cools
from the merged corona and forms stars. The remaining satellite galaxies can only
continue to form stars until their reservoirs of cold gas are exhausted. 

Luminosities for the resulting galaxies are calculated using spectral
energy distributions predicted by the stellar population synthesis models
of Bruzual and Charlot (1993, 1995).  We ignore the effects of chemical
enrichment since only spectral energy distributions for solar metallicity
are available at present. The synthesis model that we adopt is the revised
Bruzual-Charlot model, kindly provided to us by Stephane Charlot. It uses
different tracks for the late stages of stellar evolution than the earlier
version (see Charlot \and 1995).  For a single burst of star formation
with age $10$ Gyr, the revised models are redder by $\Delta (B-V) = + 0.1
$ and $\Delta (B-K) = + 0.35$ than the earlier models.  Note that subtle
differences in the choice of filter in a given band and in the type of
star used to set the zero point of a magnitude scale can lead to
differences of up to a tenth of a magnitude for a given burst age in a
particular stellar population model.  In this paper we use slightly
different filters to those employed by Cole \and (1994), in order to obtain
$B-V$ colours consistent with those derived by Charlot \and (1995) from the
same population models.

Throughout this paper, we shall use the parameters of the ``fiducial
model'' of Cole \and (1994). This is a CDM cosmology with $\Omega=1$,
$h=0.5$, $\Omega_{b} = 0.06$, where $h$ denotes the Hubble constant in
units of 100 km s$^{-1}$ Mpc$^{-1}$. The amplitude of the power spectrum of
mass fluctuations is normalised to reproduce approximately the abundance of rich galaxy
clusters by setting the {\it rms} mass fluctuation in top hat spheres of
radius $8 h^{-1}$ Mpc, $\sigma_{8}=0.67$ (White \and 1993). This fiducial
models requires strong feedback to supress star formation in dwarf
galaxies and a moderate galaxy merger rate.
	
\subsection{Bulge Formation}
\label{s:bulge}

In our extended galaxy formation scheme, all galaxies initially form stars
in a disk as hot virialised gas cools onto a collapsed dark matter halo. 
Thus we implicitly make the assumption that the timescale for star
formation is longer than the collapse timescale of the gas. A disk 
configuration is the natural outcome of the dissipative collapse of gas at 
constant angular momentum and is indeed the configuration seen in SPH 
simulations (Katz 1991, Summers, Davis \& Evrard 1993, 
Navarro, Frenk \& White 1995.) 

Spheroidal distributions of stars and galactic bulges can only form as the
outcome of merger events. In a violent merger, which we define as a merger
in which the central galaxy accretes satellite galaxies with a total mass
greater than some specified fraction of its own mass, any cold gas that is
present is turned into stars in an instantaneous burst. After the merger
and burst of star formation, if any takes place, no further stars are
added to the bulge, until the next major merger happens. Galaxies
orbiting in a common halo may continue to form stars quiescently in a disk
until their cold gas reservoir is exhausted. The central galaxy can also
form disk stars from hot gas that cools and is accreted from the hot
corona.

This scheme for bulge formation is similar to that adopted by Kauffmann \and
(1993) and  Kauffmann (1995b,c). However, as mentioned in Section
\ref{s:galform}, their galaxy formation model is different in detail to
ours. In particular, Kauffmann \and consider galaxy mergers that occur
over the lifetime of a halo as a series of binary mergers. Hence, an
individual satellite must bring in enough mass by itself in order to
satisfy the criterion for bulge formation. No account is taken of the
cumulative effects of disk heating or thickening caused by the prior
accretion of satellites that were too small on their own to cause a
significant change in the morphology of the central galaxy. In the scheme
used in this paper, we simultaneously consider all satellites that will
merge with the central galaxy over the lifetime of the halo. Hence, we
classify more merger events as `violent' mergers leading to the formation
of a bulge. Whilst neither implementation of galaxy merging is fully
accurate, the two approaches are likely to bracket what happens in reality.

To determine whether a merger is classed as `violent', we list  the
galaxies that will actually coalesce after the merger of their respective
dark matter halos. The most massive galaxy is termed the central galaxy
and the others are called its satellites. We compute the ratio of the
sum of the mass of cold gas and stars in the accreted satellites to the
mass of cold gas and stars in the central galaxy
\begin{equation}
R = \frac{ \sum_{\rm acc. \,\, sat.} M_{\rm cold \,\, gas} + M_{\rm stars}}
{(M_{\rm cold \,\, gas} + M_{\rm stars})_{\rm central \,\, galaxy}}
\end{equation}
If this ratio exceeds the specified value of a parameter, $f_{{\rm
bulge}}$, the merger is termed a `violent' merger. Following Kauffmann
\and, we set the value of the parameter $f_{{\rm bulge}}$ by
matching the morphological mix in our model to that observed locally.
(Note, however, that since galaxy mergers are treated differently in our
scheme and in that of Kauffmann \and, we have given the parameter that
defines a violent merger a different name; its numerical value will also
be different.)

In a violent merger, the disk of the central galaxy is destroyed and all
the stars present are transferred to the bulge component of the new
galaxy. All the cold gas that is present, {\it i.e.} the gas that had
cooled in individual galaxy halos prior to the merger but had not yet been
turned into stars, is then converted into bulge stars in an instantaneous
burst. If the merger is not classed as violent, the disk of the central
galaxy is preserved and no burst of star formation takes place. We have
investigated two alternatives for the fate of the stars accreted from the
satellite galaxies. In the first scheme, the stars from the satellites 
are added to the bulge of the central galaxy. In the second
scheme they are added to the disk of the central galaxy, as would happen
if the satellites are tidally disrupted and torn apart before they can reach
the core of the central galaxy. Simulations by Walker \and (1995), in
which a satellite with total mass equal to $10 \%$ of the primary galaxy
mass merges with it, reveal an intermediate situation. Roughly
$50 \%$ of the satellite survives and sinks to the core of the primary,
whilst the rest is ripped off and added to the primary's disk. We find
that there is little difference between the morphological mixes recovered
in our models using these two prescriptions and we shall adopt the first
one in the remainder of this paper.

Modern simulations of merging galaxies attempt to follow the star
formation that occurs during a merger. Mihos \& Hernquist (1994a) have
shown that a minor merger in which a disk galaxy accretes a satellite with
$10 \%$ of its own mass produce spiral arm instabilities that drive gas
into the centre of the disk and cause a burst of star formation. However,
if a bulge component is added to the primary, the spiral arms are
suppressed and a much smaller burst results. In a major merger event,
Mihos \& Hernquist (1994b) find that a burst occurs independently of the
presence of a bulge component in the galaxies, but the timing of the burst
relative to the completion of the merger and the strength of the burst do
depend upon the size of the bulge. At present numerical simulations have
explored only a small part of parameter space. Our model for bulge
formation in which a burst of star formation only occurs after a violent
merger is well motivated by the available numerical results.

After a merger event in our model, star formation takes place quiescently
in the disk of the central galaxy if a supply of cold gas is available.
The bulge-to-disk ratio is therefore a continually changing quantity. It
is quite possible for a galaxy to become a pure bulge immediately after a
violent merger and to then build up a new disk by quiescent star
formation. Thus a galaxy can move either way on the Hubble sequence,
towards early types after a violent merger in which material is added to
the bulge and towards late types during quiescent star formation when 
new stars are formed in the disk.

\subsection{Setting the parameter $f_{{\rm bulge}}$} 
\label{s:fellip}

Following Kauffmann \and (1993), we use the local morphological mix of
field galaxies to set the value of the parameter $f_{{\rm bulge}}$. Our
aim is to classify galaxies into three broad morphological classes:
disk-dominated systems, bulge-dominated systems, and systems with
intermediate bulge-to-disk ratios. We identify these broad classes with
spirals (S), ellipticals (E) and lenticulars (SO) respectively. We assign
morphological types to the model galaxies on the basis of the ratio of
bulge luminosity to disk luminosity by reference to data from Simien
\& de Vaucouleurs (1986) and de Jong (1995).

In Figure~\ref{fig:bddata} we reproduce the Simien and de Vaucouleurs
bulge-to-disk decompositions (as given in their Table 4) and de Jong's
(1995) data derived from 2D decompositions of the light in face on
spirals. A T type $T=-5$ corresponds to an elliptical, $T=-3$ to an S0
and $T=0$ to an early type spiral. There is considerable scatter in the
relationship between bulge-to-disk ratio and T type. For example, a
bulge-to-disk ratio of $\sim 1$ spans $ \sim 7 $ T types, whilst the class
$T=4$ is assigned to objects that have a spread of $10^{3}$ in
bulge-to-disk ratio. Decompositions of the light of a galaxy into bulge and
disk components are difficult to perform, and the results, for some
orientations at least, depend upon the functional form adopted for the
decomposition (de Jong 1995). Furthermore, the assignment of T-types is
fairly subjective. Naim \and (1995) find that experts can only agree on
their classifications to within 2 T-type units.  It is not clear,
therefore, whether the scatter in Figure \ref{fig:bddata} is entirely
observational or whether it reflects the fact that there is not a unique
correspondence between bulge-to-disk ratio and T-type.

Several studies in the literature have examined the morphological mix of
local field galaxies. There is a relatively small scatter amongst different
estimates of the elliptical fraction, ranging from $10 \%$ (in the Revised
Shapley Ames Catalogue, Sandage \& Tamman 1981) to $20 \%$ (in a sub-sample
of bright galaxies drawn from the RC3 catalogue by Buta \and 1994). The
scatter in the spiral fraction is larger, around $20 \%$, reflecting the
difficulty of distinguishing between spiral and S0 galaxies, particularly
at fainter magnitudes on photographic plates (Dressler 1980), when 
it becomes hard to detect the disk component in the presence of 
a significant bulge.
A recent study, based on a large homogenous sample, assigned types
to 90\% of galaxies brighter than $b_{J} = 16.44$ in the Bright APM
Galaxy Catalogue (Loveday 1995; Table~10). Incorporating the
irregular/peculiar galaxies into the spiral class and scaling up the fractions in
each type to account for unclassified objects, the morphological mix in
this catalogue is E/S0/S+Irr=13/20/67.

To fix the value of the parameter $f_{{\rm bulge}}$, we constructed mock
catalogues from the output of our models, with the same magnitude limit as
the Bright APM Galaxy Catalogue, 
taking into account any evolution in the 
galaxy properties.
Kauffmann \and classified galaxies whose bulge component is
less than $40 \%$ of the total light as spirals, galaxies whose bulge
component is $60 \% $ or more of the total light as ellipticals and
galaxies with intermediate bulge to total light ratios as S0's. If we adopt
these definitions, we obtain spiral fractions of 45\% and 55\% for $f_{{\rm
bulge}} = 0.5$ and 0.7 respectively.  Alternatively, if we allow the
maximum size of a spiral bulge to be $45 \% $ of the total light and the
minimum bulge in an elliptical to be $65 \%$ of the total light, we obtain
a morphological mix of E/S0/S of 23.5/22.5/54.0 for $f_{{\rm bulge}} =
0.5$. We shall adopt these definitions and this value of $f_{{\rm bulge}}$
throughout this paper. The corresponding cuts in bulge-to-disk ratio are
plotted in Figure \ref{fig:bddata} as solid lines. Galaxies lying below the
lower solid line are classed as spirals, galaxies above the upper solid
line as ellipticals and galaxies in between as S0s.  Although it is clear
that with some fine-tunning, we could achieve a higher spiral fraction, we
do not feel that this is justified given the large uncertainties 
surrounding morphological classification (Naim \and 1995) and the large
scatter between bulge-to-disk ratio and T-type. 

\section{Representative Examples}
\label{s:rep}

\begin{figure*}
{\epsfxsize=20.truecm \epsfysize=24.truecm 
\epsfbox[30 100 500 800]{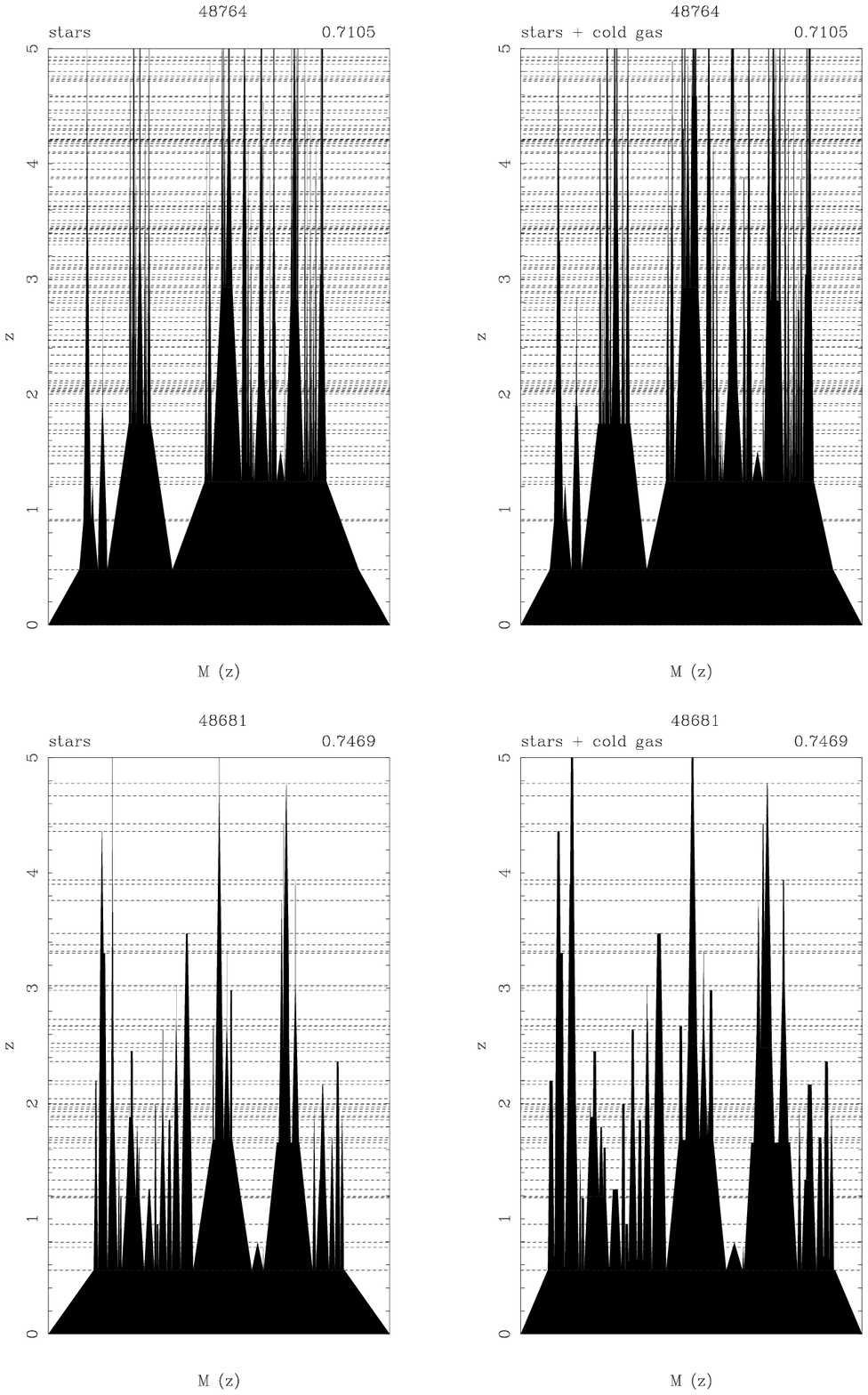}}
\caption[junk]{
Star formation histories for a cluster elliptical (upper panels) and an 
elliptical from a smaller dark matter halo 
(lower panels).
}
\label{fig:col}
\end{figure*}

\begin{figure*}
{\epsfxsize=20.truecm \epsfysize=24.truecm 
\epsfbox[30 100 500 800]{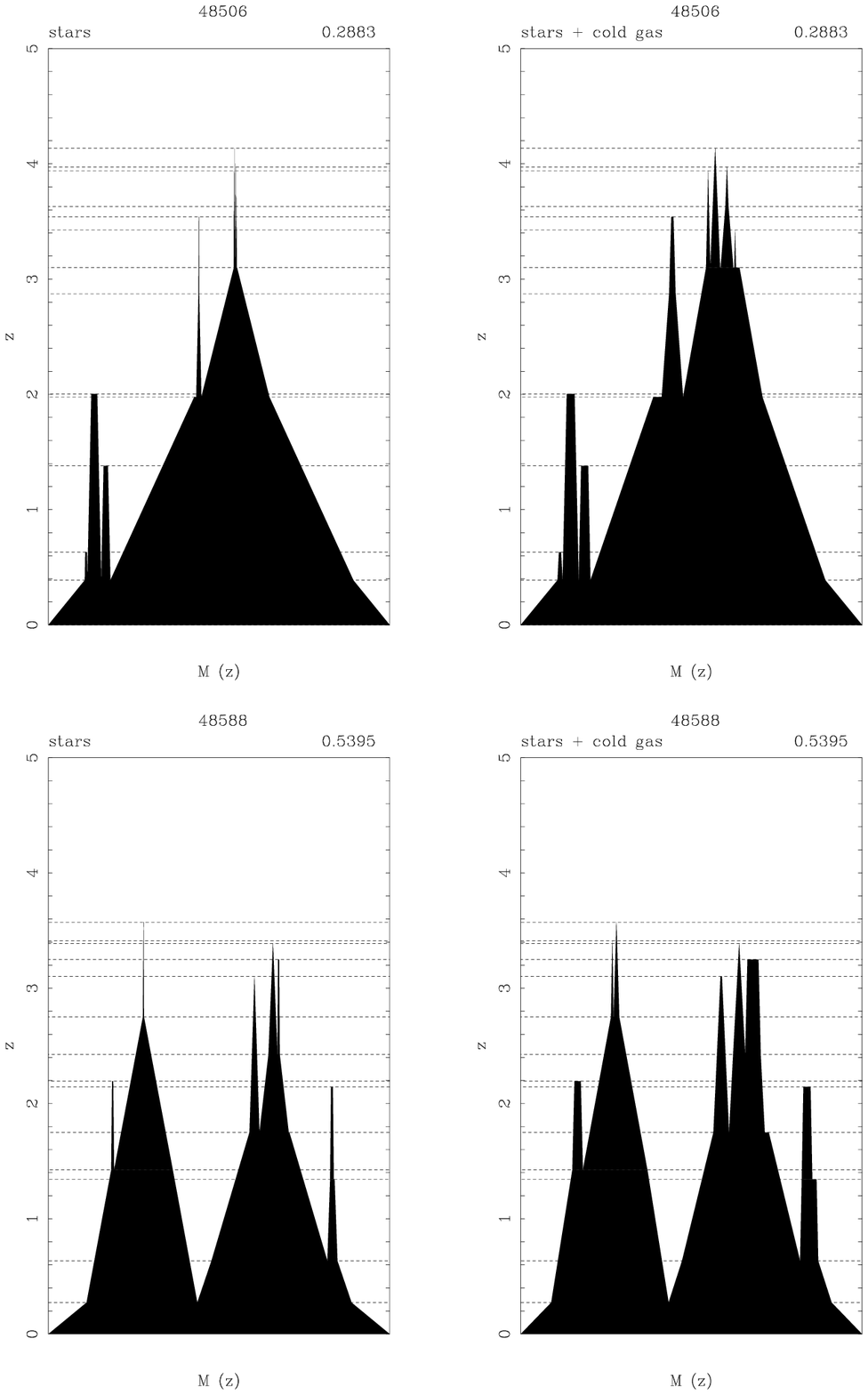}}
\caption[junk]{
Star formation histories for a spiral galaxy (upper panels) and a S0 galaxy
(lower panels).
}
\label{fig:col}
\end{figure*}

In this Section we follow the histories of a representative selection of
galaxies in our model. 
At each galaxy merger event, we have recorded properties of
the progenitor galaxies, such as their stellar mass, cold gas mass and
luminosities in various bands.
We plot the star formation history of selected galaxies in
a tree-plot, analogous to the schematic merger history of a dark matter halo
shown in Figure~6 of Lacey \& Cole (1993).  For this purpose we have traced
the history of these galaxies back to progenitor galaxies with dark matter halos of
mass $5 \times 10^{10} h^{-1} M_{\odot}$.

\begin{table*}
\begin{center}
\caption[dummy]{Representative Galaxies}
\label{tab:repgal}
\begin{tabular}{ccccccc}
\hline
\hline  
ID & $M_{\star}( 10^{11} h^{-1} M_{\odot})$ & $ L_{B} (10^{10} h L_{\odot})$
& $V_{{\rm gal}} ({\rm kms}^{-1})$ & 
B-K  & bulge/total & Type \\ \hline
48681 &  4.5    & 7.4   &   532  &   4.02  &   0.75   & E  \\
48313 &  1.8    & 3.7   &   376  &   3.68  &   0.24   & S  \\
48764 &  21.3   & 31.7  &   1016 &   4.05  &   0.71   & E  \\
48588 &  1.4    & 3.4   &   281  &   3.66  &   0.54   & S0 \\
\hline
\label{tab:prop}
\end{tabular}
\end{center}
\end{table*}

We have selected a representative range of galaxies to plot, the properties of which are 
listed in Table \ref{tab:prop}. 
For each galaxy, we show the amount of stars and  stars plus cold gas in separate 
panels as a function of redshift.
The horizontal dashed
lines mark the redshift at which a merger event took place in the history
of an object. 
The width of the black shaded regions indicates the mass of
stars or cold gas and stars that are present in a particular object.  
The mass in
cold gas and stars of the final galaxy at redshift zero is given unit width
on the x-axis. 
We know the amount of cold baryonic material in each
fragment at a merger, and, for simplicity, we perform a linear
interpolation in the tree-plots between masses at subsequent merger events
(although the actual star formation rate is exponential). 
The mass of cold
material increases between merger events as gas cools over the halo
lifetime and stars are formed.  
Hence, in a merger event, the width of the
black shaded region for a particular progenitor indicates the fraction of
the final mass of material (cold gas and stars or just stars 
depending on the plot) in the galaxy that this
fragment was responsible for bringing in. 
Termination of a shaded region at
high redshift indicates that the galaxy formed at that redshift or was
contained inside a dark matter halo with mass below the resolution limit
of the tree.

The syncronisation of merger events is an artefact of the way in which
we incorporate galaxy mergers into the block model.  
As discussed in
Section \ref{s:galform}, after the dark matter halos have merged, we
calculate a timescale for the satellite galaxies to merge with the central
galaxy of the new halo.  
Providing that this time is shorter than the
lifetime of the halo, we allow the galaxies to merge at the epoch at which
the halo is formed.  
In reality, these galaxy merger events would be spread
out in time, after the redshift at which the new halo was formed.

If we consider the tree for the spiral galaxy with ID 48506, we see that 
few mergers occured above the mass resolution of the tree.
The bulge of this object was formed by the mergers at $z=2$ and $z=0.4$.
Both these mergers are minor mergers, so there was no burst of star formation, 
and the disk of the accreting galaxy would have been preserved.
However, the accreted stars would be put into the bulge component of the 
central galaxy.

The SO tree (ID 48588) shows similarly few merger events, though the 
amount of cold gas and stars in the progenitors is more evenly 
matched in the merger at $z\sim 0.3$, which is probably a major merger.
The bulge to total ratio given in 
Table \ref{tab:prop} is measured in B-band luminosity, and so will 
underweight the bulge component relative to the disk.
Hence an object that would appear to have a high bulge to total ratio 
in terms of stellar mass, would be expected to have a lower bulge to 
total ratio in terms of B-band luminosity.

The tree plots for the ellipticals (IDs 48764 and 48681) show many more 
merger events.
In particular 48764, which is the central galaxy in a rich cluster 
at redshift zero shows many more mergers at higher $z$ than 48681, which is 
in a lower circular velocity halo.

The spiral and S0 trees look fairly similar. There are fewer merger events
than in the history of an elliptical and the last significant merger takes
place between only two progenitors.

\section{The Epoch of Bulge Formation}
\label{s:bulgeform}

\begin{figure}
{\epsfxsize=8.truecm \epsfysize=12.truecm 
\epsfbox[60 60  450 750]{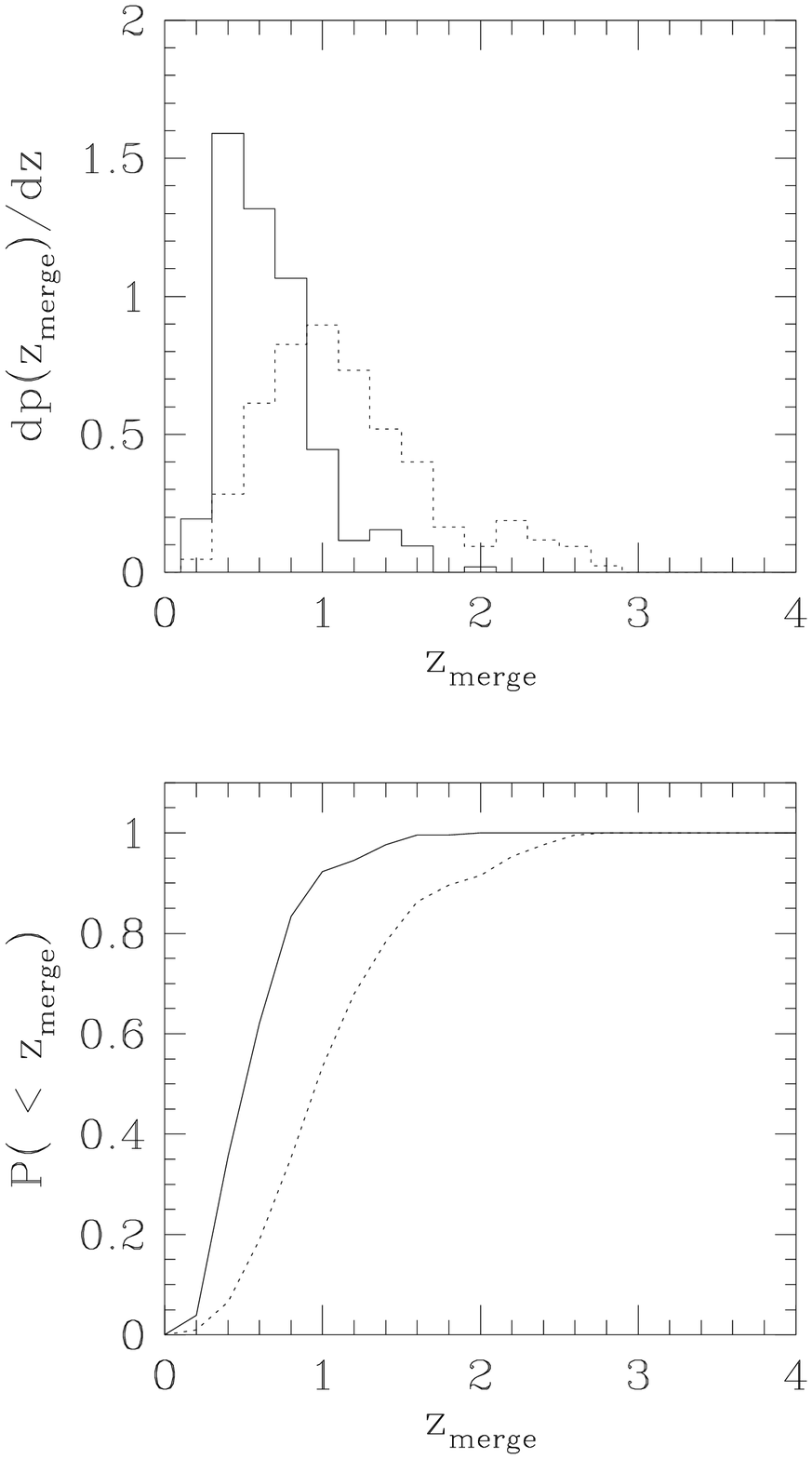}}
\caption[junk]{The distribution of the redshift of the last major merger
event for bright ellipticals (solid lines) and spirals (dashed lines). The
upper and lower panels show differential and cumulative distributions
respectively for galaxies brighter than $m_{B} - 5 \log h = -18$.  There
are 462 ellipticals that satisfy this condition and 152 spirals in our
Monte-Carlo realization.  The distributions in the upper panel have been
normalised to the total number of objects in each class.}
\label{fig:eszm}
\end{figure}

\begin{figure}
{\epsfxsize=8.truecm \epsfysize=12.truecm 
\epsfbox[60 60 450 750]{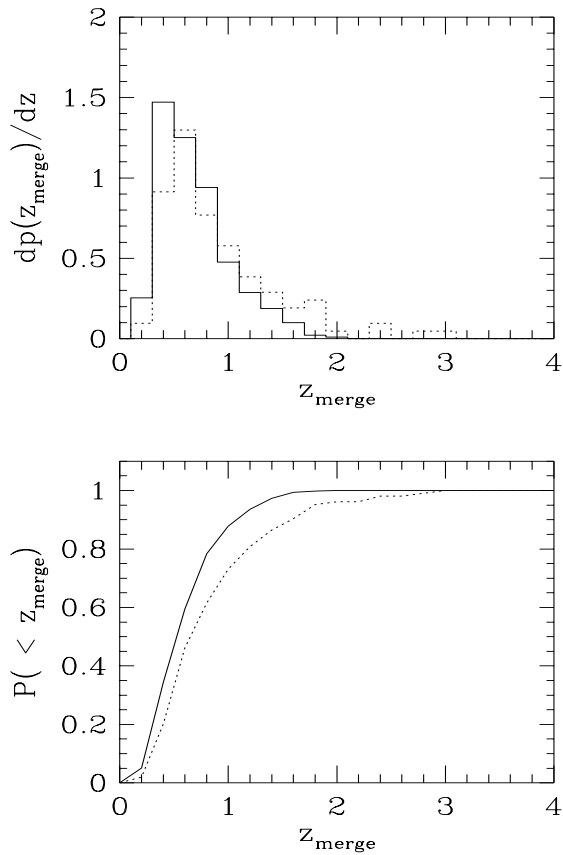}}
\caption[junk]{The distribution of the redshift of the last major merger
for bright field ellipticals (solid lines) and cluster ellipticals (dotted
lines) in halos with circular velocity $v_{c} > 1000 {\rm kms}^{-1}$. The
upper and lower panels show differential and cumulative distributions
respectively for galaxies brighter than $m_{B} - 5 \log h =
-18$.
}
\label{fig:ecfzm}
\end{figure}

\begin{figure}
{\epsfxsize=9.truecm \epsfysize=9.truecm 
\epsfbox[60 400 470 750]{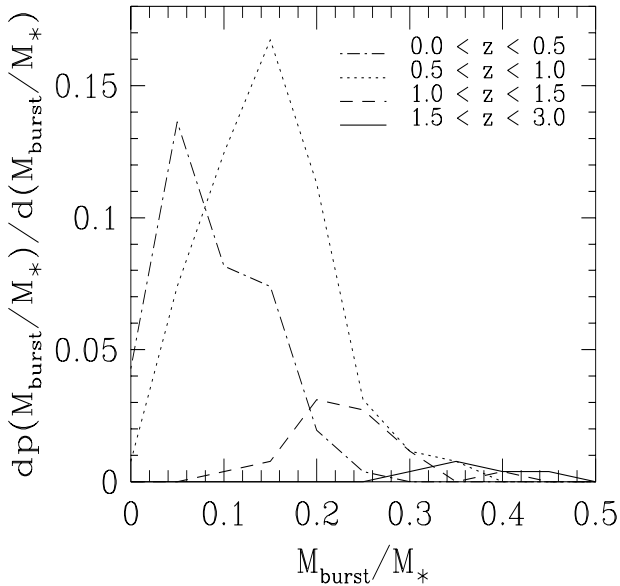}}
\caption[junk]{
The fraction of the stellar mass of a galaxy at $z=0$ that was formed in
bursts triggered by major mergers.  Galaxies classified as ellipticals at
$z=0$ on the basis their bulge-to-disk ratio and with absolute magnitudes $ m_{B}
- 5 \log h < -18 $ are considered.  The mass of stars formed in bursts is
accumulated in four redshift bins.  The curves are normalised by the total number
of galaxies that have experienced a burst.
}
\label{fig:burst}
\end{figure}

\begin{figure}
\begin{picture}(100, 600)
\put(-5,-10)
{\epsfxsize=8.truecm \epsfysize=8.truecm 
\epsfbox[50 390 460 750]{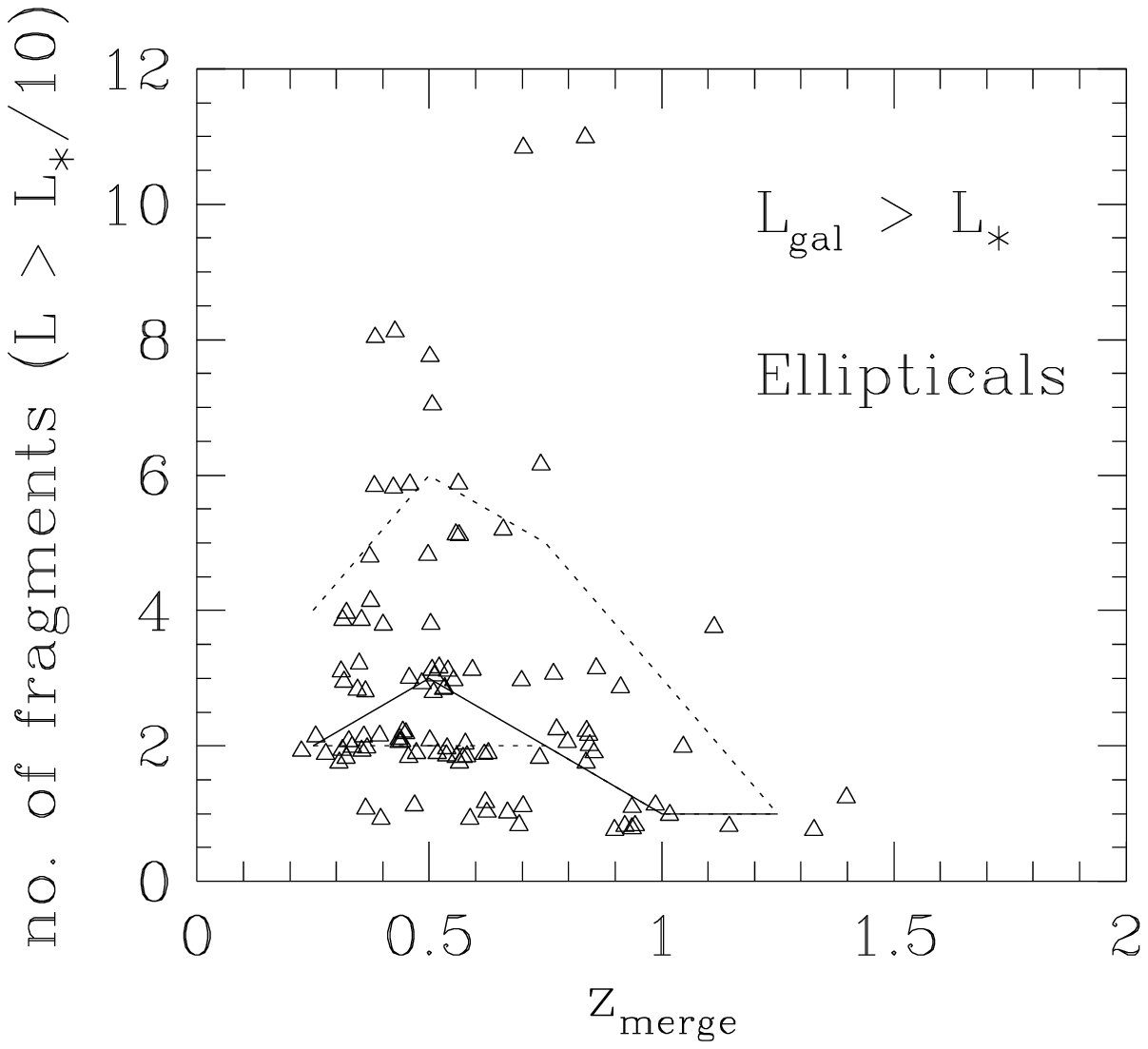}}
\put(-5,190)
{\epsfxsize=8.truecm \epsfysize=8.truecm 
\epsfbox[50 390 460 750]{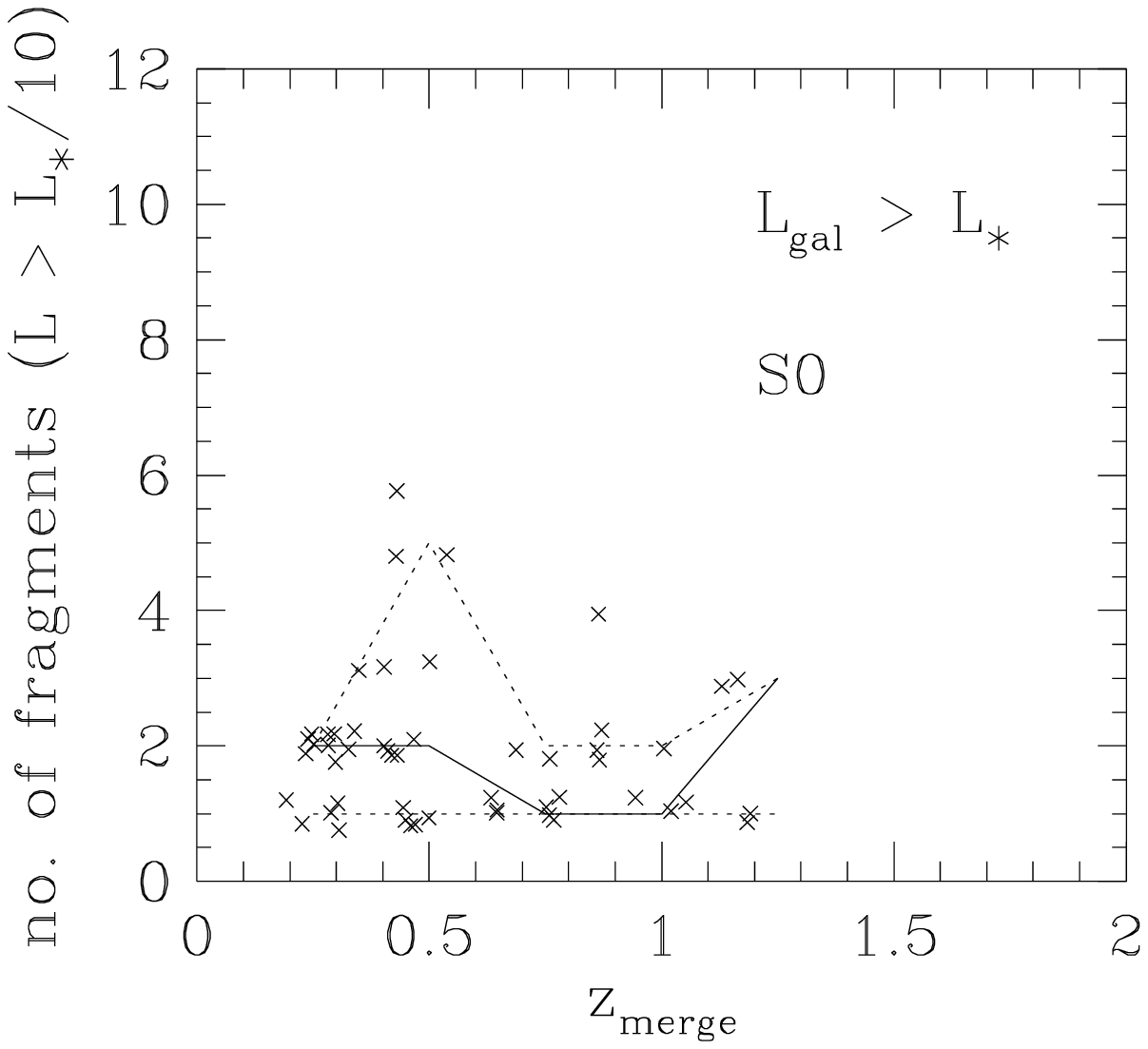}}
\put(-5,390)
{\epsfxsize=8.truecm \epsfysize=8.truecm 
\epsfbox[50 390 460 750]{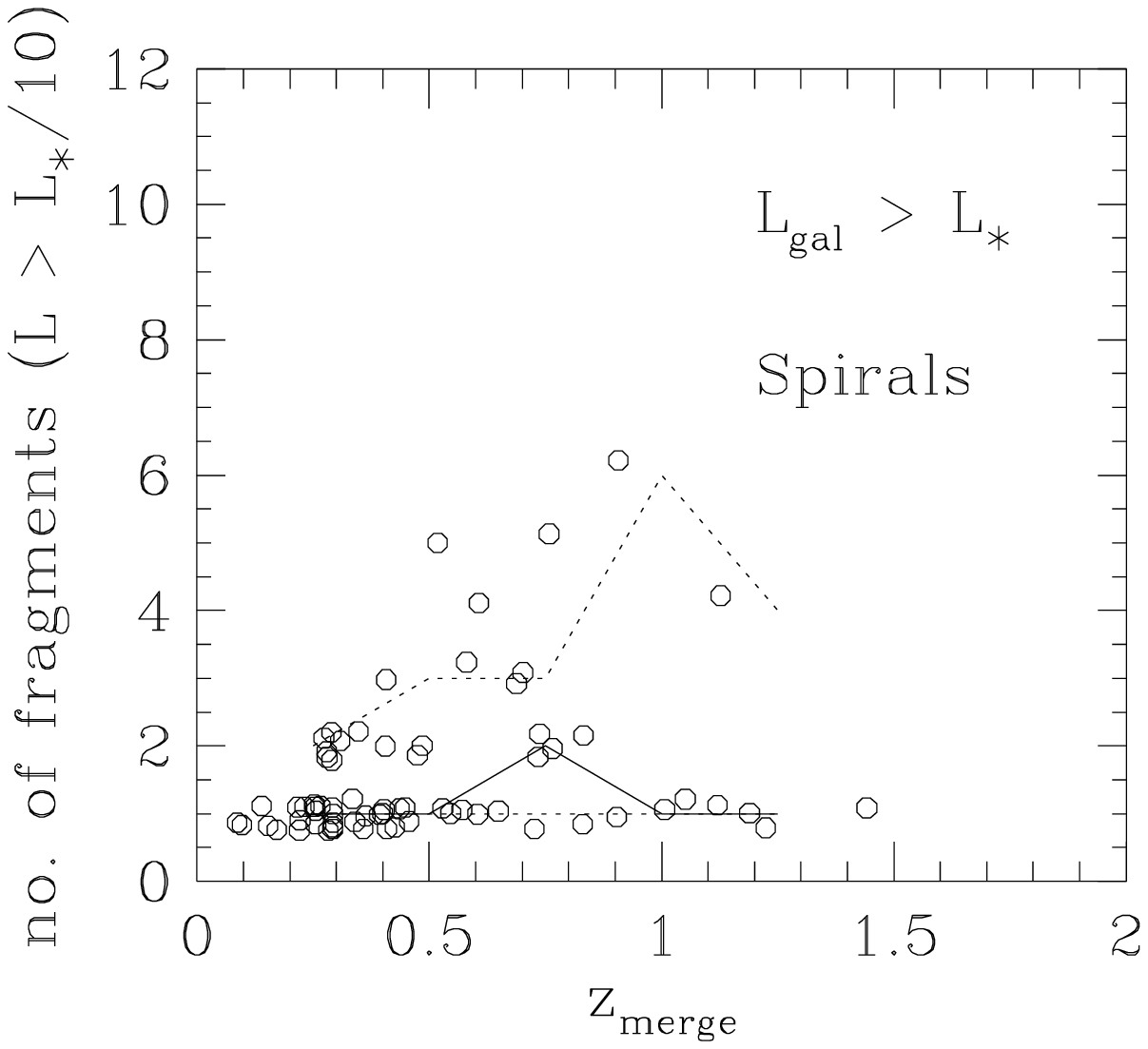}}
\end{picture}
\caption[junk]{The number of fragments, brighter than $L_{\ast}/10$, that were 
involved in the last merger event of galaxies brighter than $L_{\ast}$
today, plotted against the redshift at which the merger took place.
The dotted lines show the 20 and 80 percentiles of the distribution of 
fragments, and the solid line shows the median.
}
\label{fig:nfrag}
\end{figure}

In our model, the bulges of spiral and elliptical galaxies form by the same
mechanism, violent mergers of galaxies. It is instructive therefore to
examine whether or not there are any systematic differences between the
bulges of galaxies of different morphological type and in different
environments. 

Figure \ref{fig:eszm} shows the distribution in redshift of the last major
merger event for ellipticals (solid lines) and spirals (dotted lines). 
The average redshift of the last major merger for elliptical
galaxies is $\bar{z} = 0.68$, whilst for spirals it is $\bar{z} =
1.26$. In an $\Omega =1$ cosmology, this corresponds to a time difference
of 2 Gyr.  Our model predicts that $\sim 50 \%$ of bright ellipticals have
had a major merger between $0.0 \le z \le 0.5$, in which time only $15 \%$
of spirals have suffered a significant merger.

In Figure \ref{fig:ecfzm}, we compare the redshift of the last major merger
for ellipticals in the field and ellipticals found in halos with circular
velocity, $v_{\rm c} > 1000 {\rm kms}^{-1}$.  Cluster ellipticals form their
bulge component at a higher mean redshift, $\bar{z} = 0.94$, than field
ellipticals. This is because evolution is `accelerated' in the high density
cluster environment and so the collapse and merging of the dark matter
halos that end up in the cluster occurs at higher redshift than 
in the field.

As Figures~\ref{fig:eszm} and~\ref{fig:ecfzm} illustrate, our model
predicts that the bulges of spirals are assembled before the spheroids of
ellipticals, and that the spheroids of cluster ellipticals are assembled
before those of field ellipticals.

We have also calculated the fraction of the final stellar mass of
elliptical galaxies that is formed in bursts of star formation triggered by
major mergers throughout the history of the galaxy. Figure \ref{fig:burst}
shows the distribution of the fraction of the final stellar mass formed in
burst events, split into redshift bins.  
The curves have been normalised 
by the total number of bursts and so the relative areas under the curves 
indicate the number of bursts that take place in each redshift bin.
Every bright elliptical galaxy in our sample has experienced a major merger
between redshifts $0 < z < 3$, and $53 \%$ have had a burst in the redshift
interval $0.5 < z < 1.0$. 
The bursts in the $0.5 < z < 1.0$ interval, at a lookback
time of 6-8.5 Gyr, are, on average, responsible for $15 \%$ of the final
mass of the galaxy involved.
Bursts in the highest redshift interval, $1.5 < z < 3.0$  typically produce $\sim
30 \%$ of the final stellar mass, but occur in 
a much smaller fraction of the bright ellipticals.
This difference results from the
interplay between the mass function of fragments, their merger rate, and
the higher fraction of cold gas available at high redshift. Our results are
in accordance with recent detections of intermediate age populations in
elliptical galaxies. The data suggest that $\sim 15 \%$ of the V-band light
in ellipticals was formed in an episode of star formation approximately 5
Gyr ago, long after the formation of the bulk of the stars (Freedman 1992,
Elston \& Silva 1992).

A question that is often posed in connection with the interpretation of
faint galaxy counts is how many fragments are needed at high redshift to
make up a typical galaxy today (Guiderdoni \& Rocca-Volmerange 1988,
Broadhurst \and 1992)?  In Figure \ref{fig:nfrag} we plot the number of
fragments brighter than $L_{\ast}/10$ that participated in the last merger
event of galaxies that are brighter than $L_{\ast}$ today, against the
redshift at which the merger took place. Thus the fragments plotted are 
up to 
$\sim 2.5$ magnitudes fainter than the present-day galaxy.
Most spirals and S0s
contained only one or two such fragments at $z<1$, although a few were
`broken-up' into five or six fragments. By contrast, a significant
fraction of present-day ellipticals contained more than 4 fragments at
$z\simeq 0.5$.  These sorts of numbers are similar to those invoked by
Guiderdoni \& Rocca-Volmerange (1988) and Broadhurst \and (1992) to
account for the faint number counts. These {\it ad hoc } `merging models',
however, require further assumptions and are not directly comparable to
our hierarchical clustering scheme.

\section{Comparison with observations}
\label{s:results}

\subsection{Colour as a function of morphology, environment and redshift}

\begin{figure}
{\epsfxsize=8.truecm \epsfysize=12.truecm 
\epsfbox[60 60  450 750]{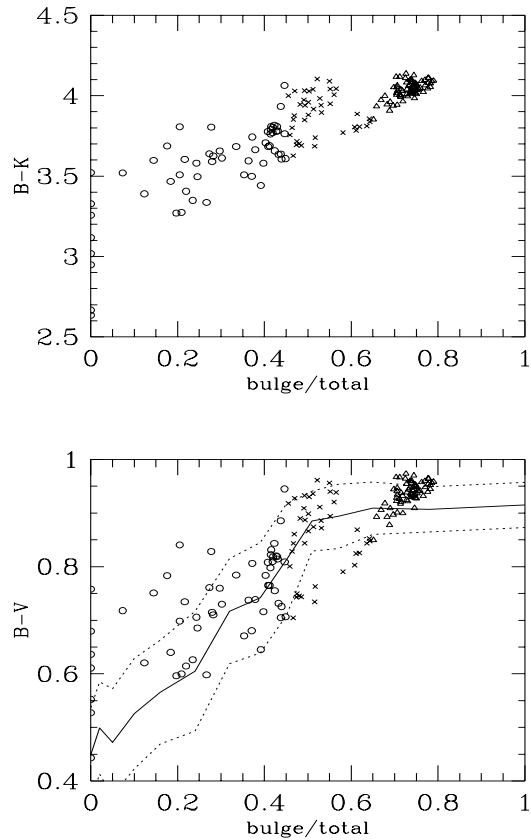}}
\caption[junk]{B-K and B-V colours as a function of bulge-to-disk ratio at
the present day.  Triangles represent ellipticals, crosses SOs and circles
spiral galaxies.  The solid line in the lower panel shows the mean B-V
colour for a sample of bright galaxies drawn from the RC3 catalogue by Buta
\and 1994; the dotted lines indicate the $1 \sigma$
scatter. Only galaxies brighter than $B=-20$ are plotted.
}
\label{fig:colbd}
\end{figure}

\begin{figure}
{\epsfxsize=8.truecm \epsfysize=12.truecm 
\epsfbox[60 60  450 750]{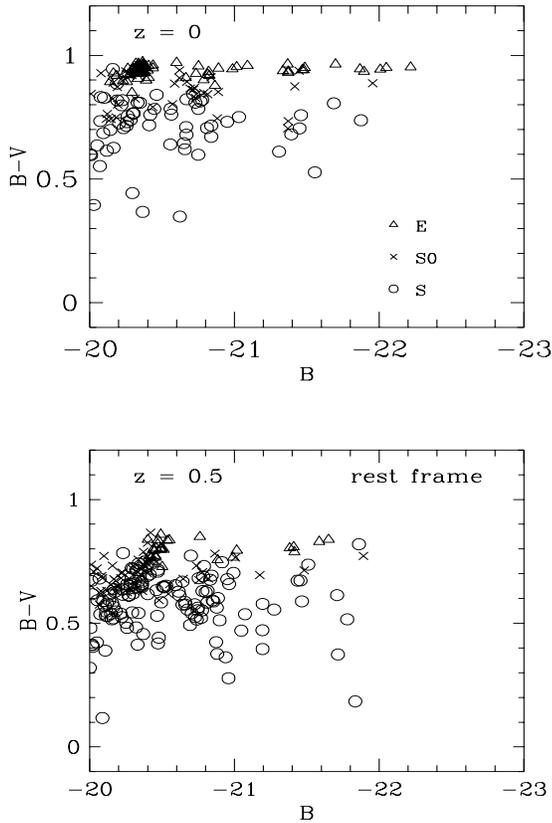}}
\caption[junk]{B-V colours as function of absolute B band magnitude. The
upper panel corresponds to $z=0$ and the lower panel to $z=0.5$. 
Triangles represent ellipticals, crosses SOs and circles spiral galaxies. 
}
\label{fig:colb}
\end{figure}

\begin{figure}
{\epsfxsize=8.truecm \epsfysize=12.truecm 
\epsfbox[60 60  450 750]{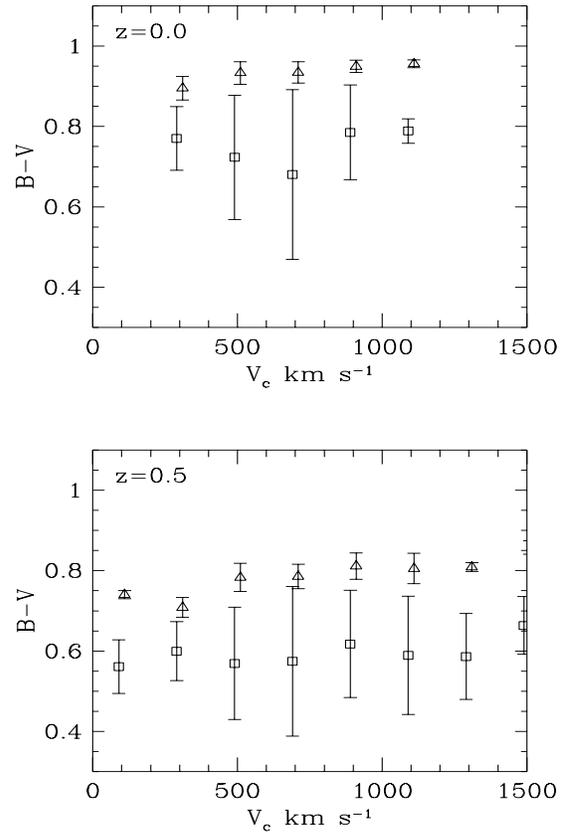}}
\caption[junk]{The mean B-V colours of elliptical and spiral galaxies as a
function of the circular velocity of the dark matter halo in which they
reside.  Triangles correspond to elliptical and squares to spiral galaxies.
The circular velocity bins are $200 \, {\rm km s}^{-1}$ wide and we have
offset the points from the centre of the bins slightly for clarity.  The
upper panel corresponds to $z=0$ and the lower panel to $z=0.5$.
}

\label{fig:bvvc}
\end{figure}

\begin{figure}
{\epsfxsize=9.truecm \epsfysize=12.truecm 
\epsfbox[100 180  550 700]{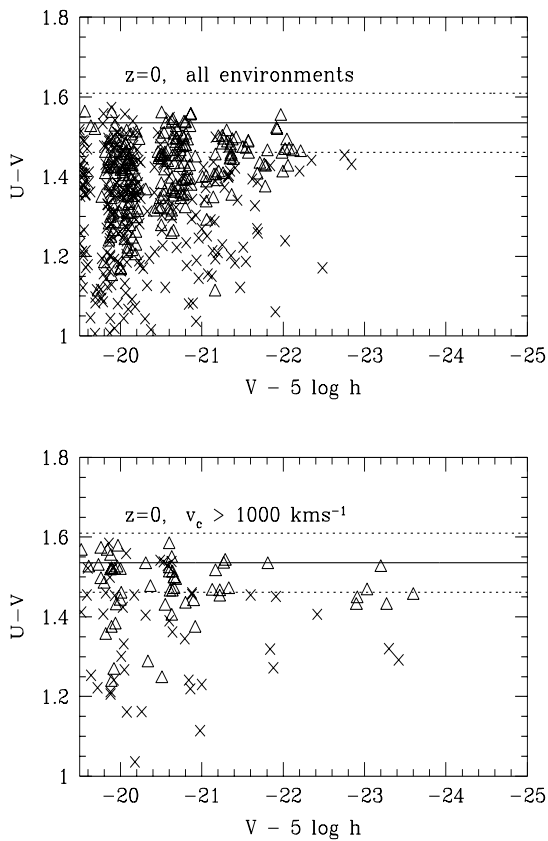}}
\caption[junk]{The colour-magnitude relation  in
different environments.
The upper panels shows elliptical (triangles) and S0 galaxies (crosses) 
found in all environments.
The lower panel shows only those galaxies found in halos with 
circular velocity $v_{\rm c} > 1000 {\rm km s}^{-1}$.
The solid lines show the median $U-V$ colour for the cluster ellipticals 
and the dotted lines show the rms scatter from our model.
These lines are reproduced in both panels for comparison.
}
\label{fig:uvvc}
\end{figure}

\begin{table*}
\begin{center}
\caption[dummy]{Elliptical galaxy colours: galaxies brighter than $B=-19$.}
\label{tab:col}
\begin{tabular}{c|cc|cc}
\hline
\hline  
colour & \multicolumn{2}{c} {cluster ellipticals} 
& \multicolumn{2}{c} {field ellipticals} \\ \hline
    & $z = 0$  & $z = 0.5$ & $z = 0$  & $z = 0.5$ \\ \hline 

B-V & $0.937 \quad (0.031) $ & $0.828 \quad (0.027) $ 
& $0.922 \quad (0.034) $ & $0.754 \quad (0.048) $ \\
B-K & $4.044 \quad (0.071) $ & $3.731 \quad (0.063) $ 
& $4.009 \quad (0.071) $ & $3.563 \quad (0.107) $ \\
U-V & $1.520 \quad (0.080) $ & $1.221 \quad (0.059) $ 
& $1.478 \quad (0.085) $ & $1.046 \quad (0.100) $ \\
\hline
\end{tabular}
\end{center}
\end{table*}

Since no quiescent star formation takes place in a bulge between merger
events, model galaxies with large bulge-to-disk ratios 
are redder than model galaxies with small bulge-to-disk ratios. 
In Figure~\ref{fig:colbd}, we plot $B-K$ and $B-V$ colours for present-day
galaxies brighter than $B=-20$, against their B-band bulge-to-disk
ratios. Different morphological types are represented with different
symbols. Also plotted is a compilation of $B-V$ colours by Buta \and (1994)
for a sample of bright galaxies taken from the Third Reference Catalogue
(RC3) (de Vaucouleurs \and 1991). Buta \and find a good correlation
between Hubble T-type and colour. Converting T-type into a bulge-to-disk
ratio (using the fit of equation~(5) of Simien and de Vaucouleurs 1986),
this correlation is shown as a solid line in the lower panel of
Figure~\ref{fig:colbd}, with the $1 \sigma$ scatter indicated
by the dotted lines. There is very good agreement between our model
predictions and the observations, although some of our galaxies with
bulge-to-disk ratios less than 20\% appear a little too red. This
discrepancy, however, is just over a tenth of a magnitude in $B-V$,
comparable to the uncertainty in colour arising from the details of the filter
choice and the template star used to set the zero point.

Figure \ref{fig:colb} shows diagrams of B-V colour {\it versus} absolute B
magnitude at redshifts $z=0$ and $z=0.5$.  As before, triangles denote
elliptical galaxies, crosses S0s and circles spiral galaxies. The
ellipticals are the reddest population and show a remarkably small spread
in colour, an issue to which we return below. Their colour-magnitude
relation is essentially flat, in apparent contradiction with the naive
expectation from hierarchical clustering models in which the largest
objects typically form last. As pointed out by Cole \and (1994), this is a
remarkable success of these models. The scatter in colour amongst spirals
is larger than amongst ellipticals, reflecting the more extended periods of
star formation in disks. S0s have intermediate colour properties. Other
than a general reddening of the populations and a slightly larger scatter in
the colours of spirals, there is little change in the appearance of the
colour-magnitude diagrams at $z=0.5$ and $z=0$. This is in spite of the
fact that a large fraction of the present-day stellar populations are still
to form at $z=0.5$.

Next, we examine the colours of bright galaxies as a function of
environment. Figure~\ref{fig:bvvc} shows the mean colours of ellipticals 
(triangles) and spirals (squares) found in halos of a given
circular velocity, at redshifts $z=0$ and $z=0.5$. The error bars give the
root mean square scatter about the mean.
There is no significant dependence of colour on halo circular velocity for
either ellipticals or spirals at the two epochs shown. For ellipticals, this
result agrees well with the data of Bower \and (1992), who found no
evidence for colour differences between galaxies in the Virgo ($v_{c} \sim
1000 \, {\rm kms^{-1}}$) and Coma ($v_{c} \sim 1400 \, {\rm kms^{-1}}$)
clusters. As this and the previous figure show, both ellipticals and
spirals redden by $\sim 0.2$ mag between $z=0.5$ and $z=0$. Photometry of
the reddest ellipticals in clusters at $z\simeq 0.5$ do indeed show that
these galaxies have slightly bluer colours than their local counterparts
(Arag\'{o}n-Salamanca \and 1993).

In Figure~\ref{fig:uvvc} we contrast properties of bright elliptical
galaxies in clusters and in the field. This figure shows $U-V$ {\it vs} $V$
colour-magnitude diagrams for the entire population of ellipticals (upper
panel) and for ellipticals found in rich clusters, which we define as halos
of circular velocity $v_{c} > 1000 {\rm km s }^{-1}$ (lower panel).
For comparison, the median colour  and root mean square scatter for the 
cluster ellipticals are reproduced in both panels. 
Ellipticals in rich clusters are predicted to
be $\sim 0.02-0.03$ mag redder than the overall elliptical population, 
reflecting the earlier mean formation epoch of galaxies
in clusters. A similar trend is seen in the observations, although the
available data are relatively poor (Larson, Tinsley and Caldwell 1981).
Table \ref{tab:col} gives the mean colours and rms scatter in brackets for the entire
population and for the subset of cluster ellipticals at redshifts $z=0$ and
$z=0.5$. Only galaxies brighter than $B=-19$ are included in this calculation, 
yielding a 
total of $210$ galaxies at $z=0$ and $100$ at $z=0.5$. The cluster
subsample comprises $45$ ellipticals at $z=0$ and $13$ at $z=0.5$.

Figure~\ref{fig:uvvc} reiterates the point made earlier regarding the 
small scatter in colour displayed by our model elliptical galaxies. This is
particularly true of ellipticals in clusters, a result first obtained by 
Kauffmann (1995c) using similar models. 
The root mean square scatter for cluster ellipticals shown by the dotted lines in 
Figure \ref{fig:uvvc} was calculated by the Median Absolute Difference 
technique used by Bower \and (1992).
The scatter we find for cluster ellipticals is $\delta (U-V) = 0.074$, 
compared with the observed value of $\delta (U-V) = 0.04$ for the 
Virgo and Coma clusters.
The scatter in our model increases for $V > -21$.
The observational data is taken from the cluster core, whereas our model 
gives the properties of all galaxies within the virial radius of the 
cluster.
Hence, a slightly bigger scatter in the colour of the cluster ellipticals 
in our model is to be expected.
Such uniformity may seem surprising in a model in which the bulk of the stars
form relatively recently and galaxy mergers are the primary mechanism for
the formation of ellipticals. The explanation for this lies in the fact
that most of the stars that end up in bright cluster ellipticals were made
in smaller fragments at higher than average redshift. When these fragments
merge, residual star formation can occur but, as discussed in Section~4,
this contributes only a relatively small fraction of the final light. Mergers
therefore mainly move galaxies along the luminosity direction, with little
change in colour. 

The scatter in the colours of cluster ellipticals is comparable at $z=0.5$
and $z=0$ (cf. Figure~\ref{fig:bvvc}). This, again, agrees surprisingly
well with the observations of Standford \and (1995) who find a similar
scatter in optical-IR colour in two Abell clusters at $z=0.374$ and
$z=0.407$ to the scatter found by Bower \and (1992) in Coma and Virgo.  

\subsection{The morphological mix in clusters and groups as a function of 
redshift}
\label{s:morden}

\begin{figure}
{\epsfxsize=9.truecm \epsfysize=9.truecm 
\epsfbox[60 400 470 750]{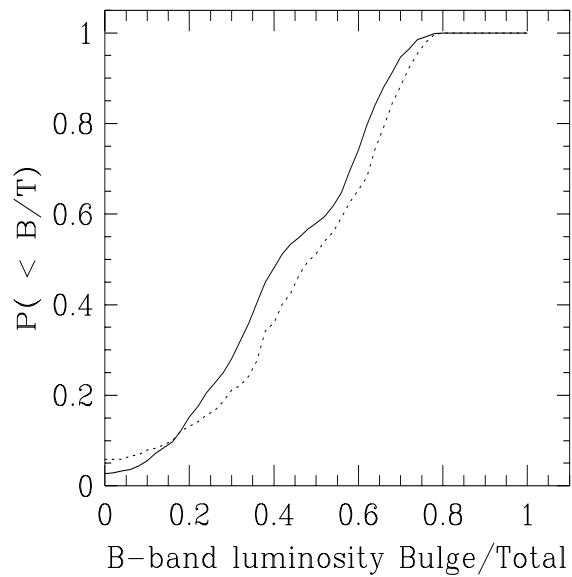}}
\caption[junk]{The cumulative distribution of objects as a 
function of bulge-to-total light ratio in the B band. The solid line
shows the whole galaxy population and the dotted line the subset of
galaxies found in clusters, ie in halos with circular velocity, 
$v_{\rm c} > \, 1000 {\rm km s}^{-1}$.
}
\label{fig:bdcum}
\end{figure}

\begin{figure}
{\epsfxsize=8.truecm \epsfysize=13.truecm 
\epsfbox[60 60  500 750]{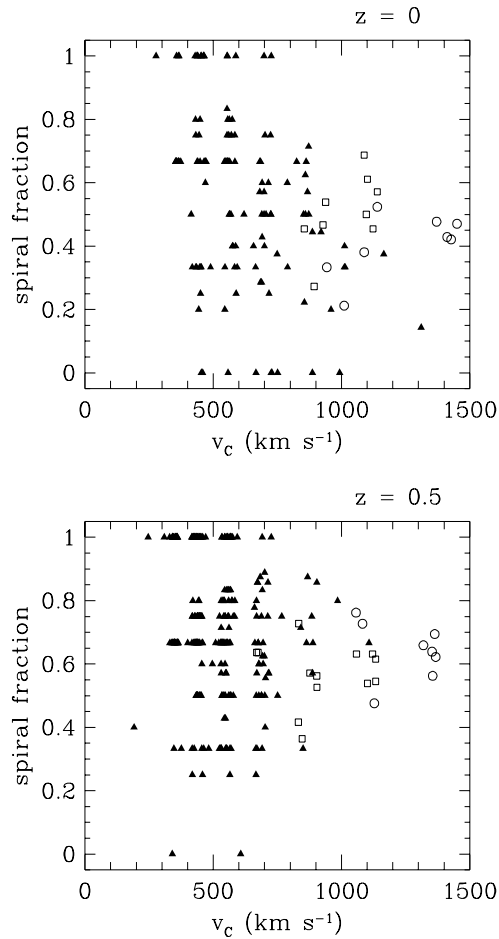}}
\caption[junk]{
The spiral fraction of galaxies in groups and clusters as a function of
halo circular velocity. The symbols indicate the richness of the group, 
defined as the number of galaxies brighter than $m_{B} - 5 \log h = -18$.
Triangles show groups with 2-10 members, squares with 10-20 members
and circles with more than 20 members. The upper panel shows model 
predictions at $z=0$ and the lower panel at $z=0.5$. 
}
\label{fig:spf_18}
\end{figure}

\begin{figure}
{\epsfxsize=8.truecm \epsfysize=13.truecm 
\epsfbox[60 60  500 750]{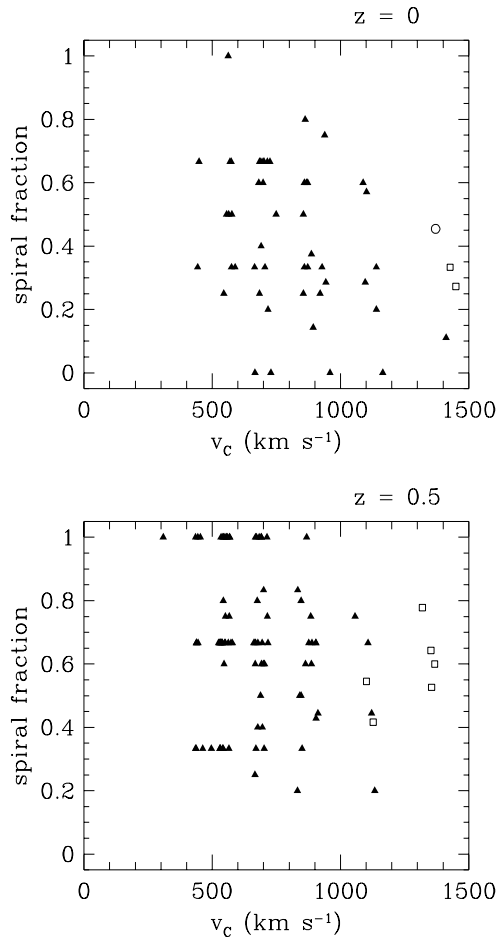}}
\caption[junk]{
As Figure~\ref{fig:spf_18} but with the group richness defined as 
the number of galaxies brighter than $m_{B} - 5 \log h = -19$.
}
\label{fig:spf_19}
\end{figure}

The best known environmental influence on galaxy morphology is Dressler's
(1980) morphology-density relation. In its simplest form this is the
statement that the fraction of ellipticals and S0s in clusters increases
rapidly with the mean projected galaxy number density.  More recently,
Whitmore and Gilmore (1991) have argued that the fundamental relation is
one between morphological mix and distance from the cluster centre.
Re-analysing Dressler's (1980) cluster sample, they found that the
elliptical fraction rises steeply within $0.5 {\rm Mpc}$ of the cluster
centre and that the fraction of S0s actually falls within $0.2 {\rm Mpc}$.
This suggests that the morphological mix depends upon a global property of
the cluster rather than on a local property such as substructure.

Since our models contain no information on the spatial distribution of
galaxies within clusters, they do not address Dressler's morphology-density
relation directly. The do, however, predict the morphological mix within
the virial radius of a cluster (defined as the radius within which the mean
density is 178 times the mean background density.) For a $10^{15} h^{-1}
M_{\odot}$ halo that collapses at the present day, this radius is $1.6
h^{-1} {\rm Mpc}$.  Examination of Figure~1 of Whitmore (1990) shows that
the elliptical fraction at this radius is $ < 20 \%$, compared with a peak value of $60 \%$
at the centre of the cluster. In Figure~\ref{fig:bdcum} we compare the
cumulative distributions of bulge-to-disk ratios for the galaxy population
as a whole (solid line) with that for galaxies located in high circular
velocity halos ($v_{\rm c} > \, 1000 {\rm km s}^{-1}$)
(dotted line). The cluster distribution is biased towards higher
bulge-to-disk ratios, implying a morphology-density relation in the same
direction as the observations. In rich clusters, $22\%$ of our model
galaxies are ellipticals, compared to 12\% in the population as a whole.

We investigate the dependence of morphology on environment further by
plotting the spiral fraction as a function of the circular velocity of the
halo to which the galaxies belong. This fraction is shown in
Figures~\ref{fig:spf_18} and~\ref{fig:spf_19} for galaxies brighter than
$M_{B} - 5 \log h = -18 $ and $-19$ respectively and spans the range of
circular velocities from poor groups to rich clusters. To obtain these
data, we constructed six realisations of the density fluctuation field,
three sampled from a Gaussian distribution in the usual way and three more
constrained to be overdense, with amplitude distributed uniformly in units
of the {\it rms} fluctuation on the scale of the block. The scatter in the
spiral fractions displayed in Figures~\ref{fig:spf_18}
and~\ref{fig:spf_19} is large for poor groups and decreases for higher
$v_{\rm c}$ clusters. In agreement with the results of Kauffmann (1995b), we find
that the richest clusters are E/S0 dominated.

We can compare the model predictions of Figures~\ref{fig:spf_18}
and~\ref{fig:spf_19} with the data obtained by Nolthenius (1993) for
groups identified in the CfA1 redshift survey. We convert circular
velocities to velocity dispersions assuming an isothermal potential and an
isotropic velocity distribution, $ \sigma_{1D} = v_{c}/ \sqrt{2}$).
Considering only galaxies brighter than $m_{B} - 5 \log h = -18$, we find
a mean velocity dispersion of  $285 \pm 72 {\rm km s}^{-1}$ for groups
with a spiral fraction greater than $50 \%$, rising to $ 352 \pm 123 {\rm
kms}^{-1}$ for groups with spiral fraction smaller than this. The
corresponding numbers for the CfA1 groups are $119 \pm 91 {\rm km s}^{-1}$
for spiral rich groups and $173 \pm 156 {\rm km s}^{-1}$ and for spiral
poor groups. Thus, the scatter in our model predictions is comparable to 
that in the CfA1 data, but the two distributions are offset from one 
another. This discrepancy is not necessarily a serious one: group velocity 
dispersions are notoriously difficult to measure from redshift surveys 
and are very sensitive to the details of group assignments (Moore, Frenk \&
White 1993). The model velocity dispersions are, of course, those 
appropriate to the dark matter halos. 

Figures~\ref{fig:spf_18} and~\ref{fig:spf_19} also show our model
predictions for the spiral fractions at redshift $z=0.5$.  For both
definitions of cluster richness, the spiral fractions in groups of all
circular velocities are larger at $z=0.5$ than at the present day. This
trend is particularly noticeable for the richest clusters. Indeed, for
$v_{c} > \, 1000 {\rm km s}^{-1}$, there are practically no clusters with a
spiral fraction smaller than 40\% at $z=0.5$, whereas at the present day
such clusters are well represented. Thus our models (like those of
Kauffmann 1995a) display a strong Butcher-Oemler effect and are consistent
with recent HST observations (Dressler \and 1994) that indicate that the
large blue fraction in high-$z$ clusters, originally discovered by Butcher
\& Oemler (1978), is, in fact, due largely to a higher fraction of bright spirals in
these clusters. Our models are also consistent with the data of
Allington-Smith \and (1993) which show that the fraction of blue galaxies
in poor groups changes much less with redshift than the corresponding
fraction in rich clusters.

Kauffmann (1995a) interpreted the origin of the Butcher-Oemler effect in
her models in terms of the different dynamical ages of clusters picked out
at different redshifts. At high redshift, a collapsed cluster of a given
mass corresponds to a rarer fluctuation than a collapsed cluster of the
same mass at low redshift. As a result, clusters seen at high redshift have
undergone more merging in the period just before their formation.
In our model this means that galaxy mergers are less likely in the 
dark matter halos that are progenitors of the cluster, because the 
halo lifetime will tend to be shorter than the galaxy merger timescale; 
at redshift zero, the dark matter halo lifetime is longer and galaxy 
mergers are more likely.
Hence, the population of galaxies that become members of a high redshift 
clusters shows a higher fraction of spirals than the population that 
become cluster members at the present day.
Furthermore, when a spiral falls 
into a cluster, its star formation continues only as long as its 
reservoir of cold gas remains. During this phase, its bulge-to-disk ratio 
will {\it decrease}. However, once the reservoir is depleted the galaxy
will redden and fade (unless it merges into the central object). In a
cluster that forms recently, there is, on average, more time for this process
to proceed than in a cluster that forms at high redshift and this results 
in a further suppression of bright spirals in low redshift clusters.

\subsection{Evolution of the Luminosity Function}

Lilly \and (1995),  Ellis (1995), Ellis \and (1995) have presented observations of the
luminosity function of galaxies in redshift bins up to $z \sim 1$. They
find little evidence for change in the luminosity function of red
galaxies, but find evolution of the blue galaxies or in the 
case of Ellis \and galaxies that have
high equivalent widths for the [O II] 3727 line, indicative of recent star
formation. Indeed for blue galaxies, Lilly \and (1995) report that the
evolution saturates at the bright end, but the luminosity function
continues to rise at fainter magnitudes.
Cole \and (1994) gave predictions for the evolution of the galaxy
luminosity function with redshift (see their Figure 19). 

In this Section we compare our model specifically with the observations of the
Canada-France Redshift Survey (CFRS) described in Lilly \and (1995), 
because we extract information on broad band colours from the 
stellar population models, rather than the equivalent widths of particular lines.
 The
CFRS sample is $I$ selected in the magnitude range $17.5 \le I_{AB} \le
22.5$. The survey consists of 730 galaxies, of which 591 have secure
redshifts with a median of $z=0.56$. Lilly \and assign a spectral type to
each galaxy by comparing  the observed $(V-I)_{AB}$ colours with the
spectral energy distributions (SED) of Coleman \and (1980). The spectral
type is then used to compute a `colour k correction' to obtain the rest
frame $B_{AB}$ and rest frame $(U-V)_{AB}$ colour. The sample is then
split into red and blue galaxies using the rest frame $(U-V)_{AB}$ colour
of an Sbc galaxy from Coleman \and (1980) as a reference point.

\begin{figure*}
{\epsfxsize=14.truecm \epsfysize=16.truecm 
\epsfbox[-30 180 540 750]{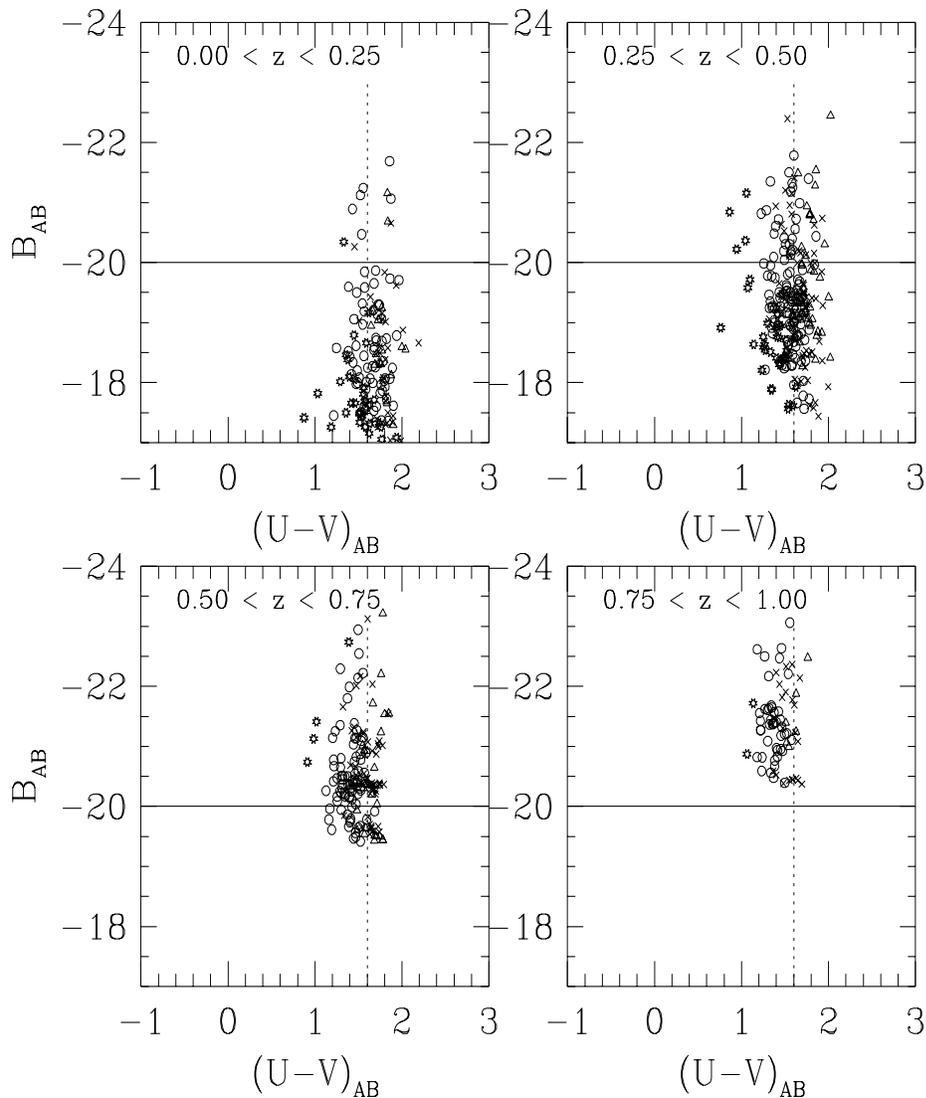}}
\caption[junk]{The colour-magnitude relation in bins of redshift for 
a sample of 730 galaxies selected from our models with magnitudes 
$ 17.5 \le I_{AB} \le 22.5$.
Triangles indicate elliptical galaxies, crosses S0's, circles 
spirals and stars objects that have a bulge component containing  
less than $5\%$ of the total light.
The vertical line shows the cut in colour that we have adopted 
to separate the galaxies into red and blue subsamples.
}
\label{fig:cfrscm}
\end{figure*}

We construct a mock CFRS catalogue from our model output by selecting
galaxies in the appropriate $I$ magnitude range. Note that we have no need
to apply $k$-corrections as we know the SED of a galaxy in our model and can
compute the rest frame colours and the appropriate transformation to the
AB magnitude system directly from this. The colour-magnitude relation that
we find in different redshift bins is shown in Figure \ref{fig:cfrscm}, which
should be compared with Figure 5 of Lilly \and (1995). For this plot we
have used 730 galaxies to match the sample size of Lilly \and . The
different symbols indicate the morphological type of the galaxies at the
redshift at which they appear in the catalogue: triangles correspond to
ellipticals, crosses to SO's, open circles to spirals and stars to objects
in which less than $5 \%$ of the total light in the B band is in the bulge
component. The dashed vertical line is at $(U-V)_{AB} = 1.60$, the rest
frame colour at which we divide the model sample into red and blue
galaxies. Lilly \and used a cut of $(U-V)_{AB}=1.3$.
The model colour-magnitude diagram is very similar to that for the CFRS data. However, the
spread of colours in the models is smaller than that observed.
These differences are not significant 
and do not affect our conclusions presented below.
The SED can change rapidly in the range of wavelengths measured by the U filter, 
being especially sensitive to assumptions made about star formation and the
choice of U filter can lead to differences in the colour.

\begin{figure*}
{\epsfxsize=16.truecm \epsfysize=20.truecm 
\epsfbox[0 10 540 750]{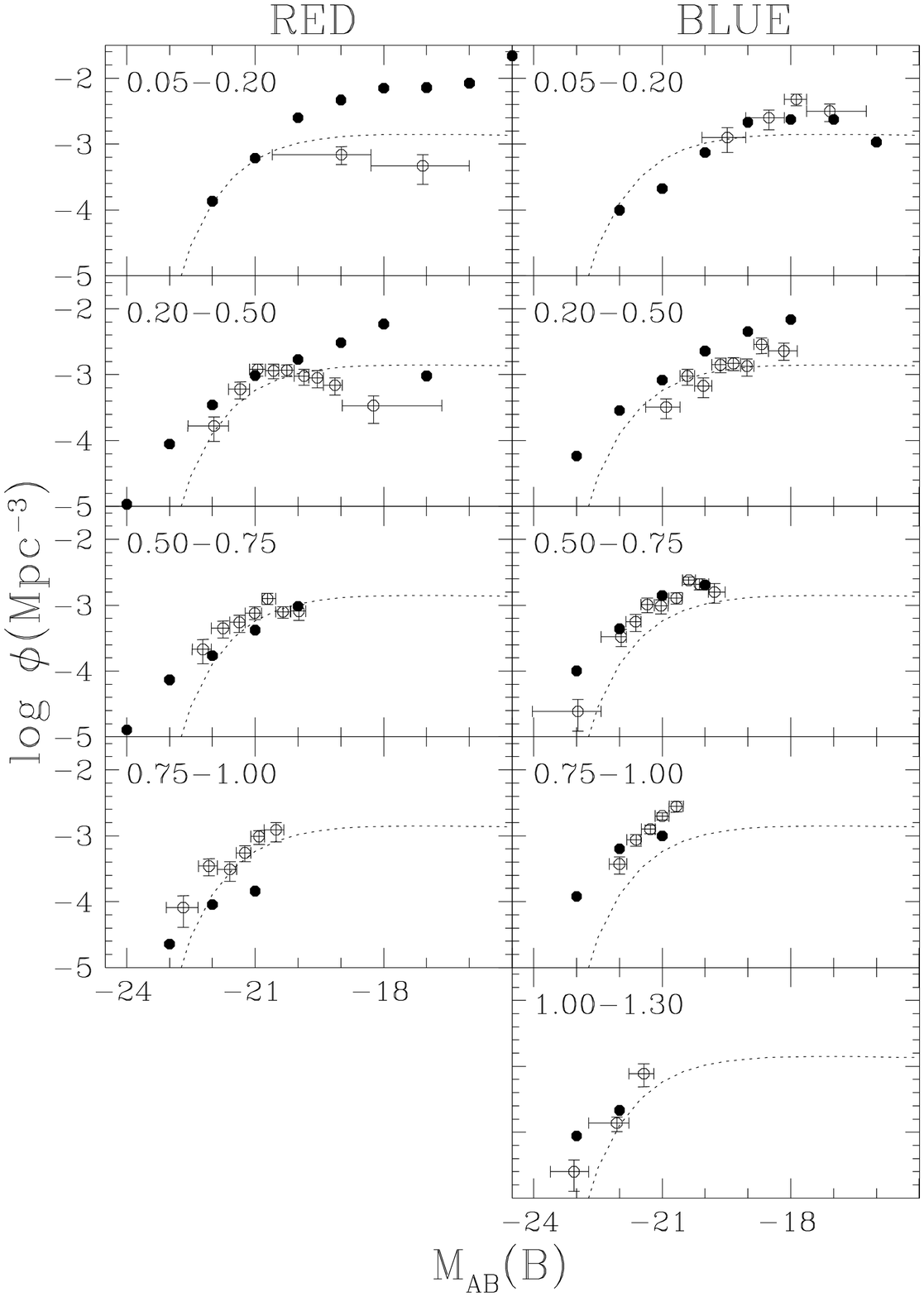}}
\caption[junk]{Luminosity functions of galaxies in our mock catalogue as 
a function of colour and redshift.
The red galaxies are shown in the left-hand panels and the blue galaxies are shown 
in the right-hand panels.
The galaxies are binned in redshift and the redshift increases down the plot, 
as indicated in the upper-left corner of each panel.
The open points with error bars show the best estimate of the CFRS luminosity 
function of Lilly \and (1995) (their Figure 3).
The filled points show the evolution of the luminosity function 
of 'red' and 'blue' galaxies in a mock CFRS catalogue constructed from our model.
The dotted line is for reference and shows the Schechter function given by 
Loveday \and (1992)
}
\label{fig:cfrslf}
\end{figure*}

Using a larger mock catalogue, containing 10000 galaxies, we compute
the luminosity function of the red and blue galaxies in redshift bins,
using the $1/V_{{\rm max}}$ formalism employed by Lilly
\and (1995). 
The luminosity functions are plotted in Figure \ref{fig:cfrslf}, 
which may be compared directly with Figure 3 of Lilly \and (1995).
The red galaxies are shown in the left column and the blue galaxies  
in the right coloumn.
Redshift increases down the plot.
The dashed line is the Schechter function
fit given by Loveday \and (1992), with parameters scaled to $AB$
magnitudes and $H_{0} = 50 {\rm km s}^{-1} {\rm Mpc}^{-1}$, and is
intended as a reference point.
The open points with errorbars are the best fit luminosity function from 
Lilly \and (1995).
The filled points show the predictions of our model.

The model luminosity function of the blue galaxies does show more evolution than
that of the red galaxies, with the luminosity function steepening,  
particularly between the $0.05 < z < 0.20$ and $0.50 < z < 0.75$ redshift 
bins.
The luminosity function of the red galaxies remains approximately 
unchanged at the bright end.
Overall, there is remarkably good agreement between our model predictions 
and the data, except at the faint end of the red galaxy luminosity function 
in the lowest redshift bin.
This discrepancy is the same one noted previously by Cole \and (1994), 
but as this plot shows, the steep faint end of the model luminosity 
function is restricted to the red galaxies.
The agreement with the data for red galaxies at high redshift is 
particularly noteworthy, since in our model $40 \%$ of the present day stars 
have still to form at $z=0.75$.

Schade \and (1995, 1996) have analysed the morphology of the galaxies in 
one of the CFRS fields.
Though the main effect responsible for the evolution of the blue luminosity 
function is the increase in the number or brightness of blue disk galaxies 
with increasing redshift, a population of CFRS objects that are bulge dominated 
but which have blue colours appears at high redshift.
At $z > 0.5$, Schade \and (1996) find that $\sim 14 \%$ of the sample, 
roughly double the fraction for  $z < 0.5$. 
are  ``blue-nucleated galaxies'' or BNG. With correction factors, 
this figure is doubled and matches the fraction of BNGs in a smaller sample 
measured with the HST.
In our mock catalogue, we can apply a similar criterion, identifying 
BNG's as galaxies that have a rest frame colour bluer than 
$ (U-V)_{AB} = 1.6 $ and a bulge to total luminosity ratio greater than 
$50 \%$.
We find that $5 \%$ of the our mock CFRS sample would be classed as 
a BNG at redshifts $z < 0.5$, and this figure rises to $22 \%$ for 
$z>0.5$.

\section{Discussion}
\label{s:discuss}

We have presented an extension of the semianalytic galaxy formation scheme
of Cole \and (1994) that allows the light of a galaxy to be split into
bulge and disk components. Galaxies are assumed to form stars quiescently
in a disk until a merger disrupts them into a spheroidal configuration.
Violent mergers are accompanied by a burst of star formation that adds
further stars to the bulge. After a major merger, a spheroid may accrete
gas from its hot corona and grow a new disk. Thus, in this scheme,
a galaxy may acquire a range of Hubble types during its lifetime. We have
presented tree-diagrams that illustrate the growth of galaxies of different
types. 

To distinguish between disks and spheroids requires adding one additional
parameter to the semianalytic model: the relative mass of a satellite
capable of disrupting a disk. We fix this, as well as the bulge-to-disk
ratios that define different morphological types, by comparing the
predicted morphological mix with that observed locally. Although for
reasons of clarity we have used the words spiral and elliptical to refer to
different classes, our simple scheme can only distinguish between broad
morphological properties based upon whether the bulge or the disk makes the
dominant contribution to the total light of the galaxy. We used this scheme
within the `fiducial' $\Omega=1$, standard cold dark matter model of Cole
\and in order to calculate the formation histories of galaxies with
different morphologies; the colours of galaxies of different types at low
and high redshift, and in field and cluster environments; the variation of
morphological mix with environment and the evolution of the luminosity
function of red and blue galaxies as a function of redshift.

We find that most spirals and S0s had only one or two progenitors with
luminosity greater than $L_*/10$ at redshift $z<1$, whereas a significant
fraction of ellipticals had five or six such progenitors. The average
redshift of the last major merger undergone by present day spirals is
$\bar{z} = 1.26$ and by present day ellipticals is $\bar{z} = 0.68$. Our
model predicts that the bulges of spirals were assembled before the
spheroids of ellipticals and the spheroids of cluster ellipticals were
assembled before those of field ellipticals. About 50\% of ellipticals, but
only about 15 \% of spirals, have undergone a major merger during the
redshift interval $0.0 \le z \le 0.5$.

Since major mergers are accompanied by a burst of star formation, we find
that about half the population of bright ellipticals has undergone a burst
of star formation in the past 6-9 Gyr. Such bursts account for around $15
\%$ of the stellar mass of the galaxy at $z=0$ and may be responsible for
the `intermediate age populations' detected in ellipticals by Rose \& Tripicco 
(1986), Freedman (1992) and Elston \& Silva (1992). They may also account
for the observations of Barger \and (1995) that suggest that $\sim 30\%$
of the ellipticals in three clusters at $z \sim 0.3$ have experienced a
burst of star formation in the 2 Gyr preceeding observation. Similarly,
some of the `E+A' spectra measured by Dressler \& Gunn (1992) in cluster
ellipticals may be the result of these merger-induced starbursts.

A striking result of our analysis, previously noted by Kauffmann (1995c), is
the flat slope and small scatter in the colour-magnitude
diagrams of cluster elliptical galaxies (see also Cole \and 1994). This is
a counter-intuitive outcome of hierarchical clustering where we might have
expected brighter galaxies to be bluer and a large scatter in colour due to
the chaotic nature of mergers. In fact, the scatter we find amongst bright
galaxies is comparable to the scatter in the observations of Bower \and
(1992) of ellipticals in the Coma and Virgo clusters, 
when one takes into account that the observations are of the cluster 
cores, whereas our models cannot revolve within the virial radius. A flat
colour-magnitude relation in the models results from the fact that the bulk
of the stars that end up in ellipticals formed at high redshift even though
the galaxies themselves were not assembled until much later. The small
scatter results, in part, from the flatness of this relation. The observed
colour-magnitude diagram of cluster ellipticals is, of course, not
completely flat, but has a small gradient of increasing redness with
increasing brightness. In the context or our models this gradient must be
due to metallicity effects which, in the absence of the appropriate stellar
population synthesis models, we have been forced to neglect. We plan to
study these effects in a forthcoming paper using Worthey's (1994) new,
metallicity dependent synthesis models. Whether our models will be able to
reproduce the observed gradient whilst retaining a small scatter in colour
remains an open question.

An important prediction of our models is the similarity between the
colour-magnitude diagrams at redshifts $\sim 0.5$ and at the present
day. The main evolutionary effect is a blueing of the galaxies by about
0.15 mag in B-V, similar to that observed in cluster
ellipticals by Arag\'on-Salamanca \and (1993). This results from a
combination of passive evolution of the stellar populations and evolution
of the star formation rate. Other than that, our models of elliptical
galaxies predict almost no increase in the scatter in colour at a given
luminosity out to $z\simeq 0.5$. For spirals, there is a moderate increase
in the scatter at high redshift. Both at $z=0.5$ and at the present, the
mean B-V colours of ellipticals and spirals in clusters are almost
independent of cluster richness.

Rich clusters in our model contain a  different mix of morphological 
types if they collapse at redshift zero, compared with clusters that 
collapse at $z=0.5$.
Kauffmann (1995a) has demonstrated that a cluster of a given mass at high 
redshift is assembled more rapidly than a cluster of the same 
mass at the present day.
In the context of our model, this means that the lifetime of a dark matter 
halo that is a progenitor of a cluster at high redshift will typically be much 
shorter than the timescale for galaxy mergers.
Hence the population of galaxies that become members of high redshift 
clusters shows a higher fraction of spirals, in agreement with HST 
observations of $z \sim 0.4$ clusters (Dressler \and 1994).
The dark matter halo lifetime for low redshift cluster progenitors is 
longer, and so galaxy mergers are more likely.
Hence in our model rich clusters that form at low redshift are dominated by 
E/S0 galaxies, reproducing the morphology-density relation discovered in 
the real universe by Dressler (1980).

We have compared the predictions of our model with observations of
the luminosity function at different redshifts.
When the model galaxies are separated by colour, the luminosity function of 
the blue galaxies evolves in steepness and brightness between 
$z \sim 0.1 $ and $z \sim 0.6$, remaining fairly constant up to $z \sim 1$.
The luminosity function of the red galaxies stays approximately constant with
redshift.
At higher redshifts, we do find objects that have blue colours and which 
are bulge dominated. 
This is expected if these galaxies have recently experienced a merger event, 
with the blue colours due to a burst having taken place in the 
case of a violent merger or because stars that had recently formed in a disk 
have now been incoporated into the bulge for the case of a minor merger.

The calculations presented in this paper serve to demonstrate that the
simplest possible prescription for the distinction between disks and
spheroids that is compatible with hierarchical clustering goes a long way
towards explaining many of the systematic trends observed in the galaxy
population. Our models are necessarily idealised and ignore processes such
as internal galactic phenomena or dynamical friction in clusters, which
have almost certainly helped to shape the complex morphologies of real
galaxies. It is therefore quite remarkable that they are able to account,
at least in an approximate way, for such fundamental properties as the
morphology-density relation, the colour-magnitude diagram, the
Butcher-Oemler effect and the evolution of the luminosity function of
galaxies of different colours. A number of critical observations such as
measurement of colour-magnitude diagrams of cluster galaxies at high
redshift will help to test this paradigm. 

\section*{Acknowledgements}

We would like to thank Stephane Charlot for providing us with a revised
stellar population model. Richard Nolthenius kindly provided his data in
electronic form. We thank Richard Bower for 
his detailed comments on an earlier version of this paper 
and for the loan of statistical software.
We acknowledge useful conversations with
Guinivere Kauffmann, Gary Mamon and Julio Navarro. CMB acknowledges a
PPARC research assistantship and SMC a PPARC Advanced Fellowship. This 
work was supported in part by a PPARC Rolling Grant. 

\section*{References}

\setlength{\parindent}{0cm}

\def\refe {\par \hangindent=1cm \hangafter=1 \noindent}
\def\aj { Astron. J., \rm}
\def\apj { Astroph. J., \rm }
\def\apjs { Astroph. J. Suppl., \rm }
\def\mn { MNRAS, }
\def\apl { Ap. J. (Letters), }

\refe Allington-Smith, J.R., Ellis, R.S., Zirbel, E.L., Oemler, A., 1993, \apj 
      404, 521
\refe Arag\'{o}n-Salamanca, A., Ellis, R.S., Couch, W.J., Carter, D., 1993, \mn 
      262, 764
\refe Barger, A.J., Arag\'{o}n-Salamanca, A., Ellis, R.S., Couch, W.J., Smail, 
      I., Sharples, R.M., 1995, \mn in press 
\refe Barnes, J.E., Hernquist, L., 1992, Ann. Rev. Astron. Astroph., 30, 705
\refe Bond, J.R., Cole, S., Efstathiou, G., Kaiser, N., 1991, \apj 379, 440
\refe Bower, R.G., 1991, \mn 248, 332
\refe Bower, R.G., Lucey, J.R., Ellis, R.S., 1992, \mn 254, 589
\refe Bower, R.G., 1995 astro-ph/9511058
\refe Broadhurst, T.J., Ellis, R.S., Glazebrook, K., 1992, Nature, 355, 55
\refe Bruzual, G., Charlot, S., 1993, \apj 405, 538
\refe Buta, R., Mitra, S., de Vaucouleurs, G., Corwin, H.G., 1994, \aj 107, 118
\refe Butcher, H.R., Oemler, A., 1978, \apj 219, 18
\refe Butcher, H.R., Oemler, A., 1984, \apj 285, 426
\refe Charlot, S., Worthey, G., Bressan, A., 1995, \apj in press
\refe Cole, S., 1991, \apj 367, 45
\refe Cole, S., Kaiser, N., 1988, \mn 233, 637
\refe Cole, S., Arag\'{o}n-Salamanca, A., Frenk, C.S., Navarro, J.F., Zepf, S.E., 1994, 
      \mn 271, 781
\refe Coleman , G.D., Wu, C.C., Weedman, D.W., 1980 \apj Suppl. 43, 393
\refe Colless, M., Schade, D., Broadhurst, T.J., Ellis, R.S., 1994, \mn 267, 1108 
\refe Couch, W.J., Ellis, R.S., Sharples, R.M., Smail, I., 1994, \apj 430, 121
\refe Dressler, A., 1980 \apj 236 351
\refe Dressler, A., Gunn, 1992, \apjs 78, 1
\refe Dressler, A., Oemler, A., Sparks, W.B., Lucas, R.A., 1994, \apj 435, L23 
\refe Eggen, O.J., Lynden-Bell, D., Sandage, A.R., 1962, \apj 136, 748
\refe Ellis, R.S., 1995, astro-ph/9508044
\refe Ellis, R.S., Colless, M., Broadhurst, T., Heyl, J., Glazebrook, K., 1995, MNRAS, 
      in press
\refe Elston, R., Silva, D., 1992, \aj 104, 1360
\refe Farouki, R., Shapiro, S.L., 1981, \apj 243, 32
\refe Frenk, C.S., Baugh, C.M., Cole, S., 1995 to appear in proccedings 
      IAU symp. 171, New light on Galaxy Evolution, eds. Bender, R., 
      Davies, R.L., (Kluwer)
\refe Freedman, W., 1992, \aj 104, 1349
\refe Guiderdoni, B., Rocca-Volmerange, B., 1988, Astron. \& Astroph., 205, 369 
\refe Gott, J.R., Thaun, T.X., 1976, \apj 204, 649
\refe Gunn, J.E., Gott, J.R., 1972, \apj 176, 1
\refe Heyl, J.S., Cole, S., Frenk, C.S., Navarro, J.F., 1995, \mn 274, 755
\refe de Jong, R.S., 1995 PhD Thesis Groningen
\refe Joseph, R.D., 1990, in Dynamics and Interactions of Galaxies, ed 
      Weilen, R., (New York:Springer) 132
\refe Katz, N., 1991, \apj 368, 325
\refe Katz, N., Hernquist, L., Weinberg, D.H., 1992, \apj 399, L109
\refe Kauffmann, G., 1995a, \mn 274, 153
\refe Kauffmann, G., 1995b, \mn 274, 161
\refe Kauffmann, G., 1995c, astro-ph/9502096
\refe Kauffmann, G., Guiderdoni, B., White, S.D.M., 1994 \mn 267,  981 
\refe Kauffmann, G., White, S.D.M., Guiderdoni, B., 1993, \mn 264, 201
\refe Lacey, C.., Silk, J., 1991, \apj 381 14
\refe Lacey, C., Cole, S., 1993, \mn 262, 627
\refe Lacey, C.., Guiderdoni, B., Rocca-Volmerange, B., Silk, J., 1993, 402, 15
\refe Larsen, R.B., 1975, \mn 173, 671
\refe Larson, R.B., Tinsley, B.M., Caldwell, C.N., 1980, \apj 237, 692
\refe Lilly, S.J., Tress, L., Hammer, F., Crampton, D., Le F\`{e}vre, O., 1995, 
      \apj in press
\refe Loveday, J., 1996, MNRAS, 278, 1025
\refe Loveday, J., Maddox, S.J., Efstathiou, G., Peterson, B.A., 1995, \apj 442, 457
\refe Loveday, J., Peterson, B.A., Efstathiou, G., Maddox, S.J., 1992, \apj 390, 338
\refe McGaugh, S., 1994, Nature, 367, 538
\refe Marzke, R.O., Huchra, J.P., Geller, M.J., 1994, \apj 428, 43
\refe Mihos, J.C., Hernquist, L., 1994a, \apj 425, L13
\refe Mihos, J.C., Hernquist, L., 1994b, \apj 431, L9
\refe Moore, B., Frenk, C.S., White, S.D.M., 1993, \mn 261, 827
\refe Moore, B., Katz, N., Lake, G., Dressler, A., Oemler, A., 1995, Nature, 
      in press
\refe Naim, A., Lahav, O., Buta, R.J., Corwin, H.G., de Vaucouleurs, G., 
      Dressler, A., Huchra, J.P., van den Bergh, S., Raychaudhury, S., 
      Sodre, L., Storrie-Lombardi, M.C., 1995, \mn 274, 1107
\refe Navarro, J.F., Frenk, C.S., White, S.D.M., 1995, \mn 275, 56
\refe Navarro, J.F., White, S.D.M., 1993, \mn 265, 271
\refe Negroponte, J., White, S.D.M., 1983, \mn 205, 1009
\refe Nolthenius, R., 1993, \apjs 85, 1
\refe Ostriker, J.P., 1980, Comm. Astrophysics, 8, 177
\refe Press, W.H., Schechter, P.L., 1974, \apj 187, 425
\refe Rose, J.A., Tripicco, M.J., 1986, \aj 92, 610
\refe Sandage, A., Freeman, K.C., Stokes, N.R., 1970, \apj 160, 831
\refe Schade, D., Lilly, S.J., Crampton, D., Hammer, F., Le F\`{e}vre, O., Tresse, L., 
      1995, \apj 451, L1
\refe Schade, D., Lilly, S.J., Le F\`{e}vre, O., Hammer, F., Crampton, D., 1996, \apj 
      in press
\refe Schweizer, F., Seitzer, P., 1992, \aj 104, 1039
\refe Simien, F., de Vaucouleurs, G., 1986, \apj 302, 564
\refe Stanford, S.A., Eisenhardt, P.R.M., Dickinson, M., 1995, \apj 450, 512
\refe Summers, F.J., Davis, M., Evrard, A., 1995, \apj 454, 1
\refe Toomre, A., 1977 {\it The Evolution of Galaxies 
      and Stellar Populations}, 
      ed. B.M. Tinsley, R.B. Larsen, New Haven: Yale Univ. Press
\refe Walker, I.R., Mihos, J.C., Hernquist, L., 1995, astro-ph/9510052
\refe White, S.D.M., Efstathiou, G., Frenk, C.S., 1993, \mn 262, 1023
\refe White, S.D.M., Frenk, C.S., 1991, \apj 379, 52
\refe Whitmore, B.C., Gilmore, D.M., 1991, \apj, 367, 64
\refe Whitmore, B.C., 1990 in Clusters of Galaxies eds. Oegerle, W.R., 
      Fitchett, M.J., Danly, L., 139
\refe Worthey, G., 1994, \apjs 95, 107

\end{document}